\documentclass[a4paper,11pt]{article}
\pdfoutput=1 

\usepackage{jheppub} 

\usepackage[T1]{fontenc} 
\usepackage{amsmath}
\usepackage{amssymb}
\usepackage{ytableau}
\usepackage{mathrsfs}
\usepackage{multirow}
\title{\boldmath Correlators of Mixed Symmetry Operators in Defect CFTs}

\author[]{Sunny Guha and}
\author[]{Balakrishnan Nagaraj}

\affiliation[]{George P. and Cynthia Woods Mitchell Institute for Fundamental Physics and Astronomy,
\\Texas A\&M University,
\\ College Station, TX 77843, USA}

\emailAdd{sunnyguha@physics.tamu.edu}
\emailAdd{nbala@physics.tamu.edu}

\abstract{We use the embedding formalism technique to study correlation functions of a $d$-dimensional Euclidean CFT in the presence of a $q$ co-dimensional defect. The defect breaks the global conformal group $SO(d+1,1)$ into $SO(d-q+1,1)\times SO(q)$. We calculate all possible invariant structures that can appear in one-point, two-point and three-point correlation functions of bulk and defect operators in mixed symmetry representation. Their generalization to $n$-point correlation functions are also worked out. Correlation functions in the presence of a defect, in arbitrary representation of $SO(q)$, are also calculated.}

\begin{document} 
\begin{flushright}
{
{\sc MI-TH-1883}\\
}
\end{flushright}
\maketitle
\flushbottom

\section{Introduction} \label{sec:intro}
A study of Conformal Field Theories (CFT) is essential for a better understanding of various phenomena that involve phase transitions, critical points, AdS/CFT duality etc.  CFTs are also important from the point of view that all Quantum Field Theories flow under renormalization towards scale invariant fixed points, where the scale invariance is often enhanced to a larger conformal symmetry. Thus a classification of CFTs is vital for our understanding of Quantum Field Theories. The local data of a unitary CFT are specified by the spectrum of operators together with the three-point coefficients. Solving a CFT generally refers to finding this data. Due to the richness of the Virasoro algebra (extension of global conformal group) in 2-dimensions, some CFT models (e.g. Minimal models) have been solved exactly. The most efficient way to extract information about the CFT data in higher dimensions is to use the crossing relations. This has given rise to the bootstrap program \cite{firstboot,carving,tasi,bootrev}.

Conformal theories with defects have a range of applications from condensed-matter physics to particle physics. Experimental systems inherently contain a boundary (a type of defect) making the study of defects essential. The simplest example of a defect is a co-dimension one defect, a boundary. Boundary defects (within the context of CFT) in 2 dimensions have been thoroughly studied in \cite{cardy1,cardy2}. Boundary defects in general dimensions were first studied beginning in \cite{osborn} and an embedding formalism was set up for co-dimension one defects in \cite{boundaryboots}. The extension to general co-dimension defects was studied in \cite{marco,gadde}. 

A Euclidean CFT with defects has both bulk operators and defect local operators (which reside on the defect). The defect local operators transform under the broken conformal group $SO(p+1,1)\times SO(q)$ where  $p+q=d$ ($q$ is the co-dimension of the defect). In addition to the CFT data of the bulk sector, there is also the CFT data of the defect sector and the couplings between the two sectors. In this work we will refer to the entire theory with both the sectors as a \textit{defect CFT}. The presence of a defect induces a rich structure in the bulk sector. For example, a bulk local operator ($O$) near a defect can be expanded in terms of defect local operators ($\hat{O}$),
\begin{equation} \label{defect expansion}
    O(x^{\mu}) \sim \sum_k b_{O\hat{O}_k}\frac{\hat{O}_k(x^a)}{|x^i|^{\Delta-\hat{\Delta}}} + \dots  ,
\end{equation}
where $x^a$ and $x^i$ are coordinates parallel and perpendicular to the defect respectively. The decomposition ($\ref{defect expansion}$) leads to bulk local operators having non-zero vacuum expectation values. We can also expand two defect operators in terms of other defect operators (regular Operator Product Expansion(OPE)),
\begin{equation}
    \hat{O}_1(x^a)\hat{O}_2(y^a) \sim \sum_k \hat{f}_{12k} \frac{\hat{O}_k(y^a)}{|x^a-y^a|^{\Delta_1+\Delta_2-\hat{\Delta}_k}}+\dots \: .
\end{equation}
The defect sector behaves like an ordinary $p$-dimensional CFT with $SO(p+1,1)$ as its conformal group and an additional $SO(q)$ global symmetry. Since the defect sector is exchanging energy with the bulk there is no conserved stress-energy tensor for the theory living on the defect \cite{McAvity:1993ue}.

Crossing symmetry relations constrain the data of a CFT. These equations can be solved numerically (e.g. \cite{firstboot,carving}) or analytically (e.g. \cite{analboot}). An ordinary CFT gives rise to a crossing relation at the four-point correlator level. However a defect CFT gives rise to crossing relations starting at the two-point correlator level. The knowledge of correlation functions (tensor structures) is essential in the study of crossing relations. In \cite{marco} tensor structures for symmetric traceless operators were found for two-point correlators. In this work we build upon those results and extend it to $n$-point correlators of operators in arbitrary mixed symmetry representations. In particular, we compute all possible invariants and tensor structures that could arise in a one-point, two-point and three-point correlator of various bulk and defect operators. We also indicate the invariants that could arise in an $n$-point correlator. One and two-point correlators for defects in arbitrary representations of $SO(q)$ are also computed.

The structure of this paper is as follows: In section 2, we lay out the mechanics of encoding mixed symmetry tensors of CFTs involving defects in terms of polynomials in embedding space. We briefly discuss the two possible types of conformal defects in section 3. In section 4, we consider the one-point functions of bulk operators and comment on some non-zero cases. We consider several examples and illustrate the duality between co-dimension $q$ defect and co-dimension $d+2-q$ defect. In section 5 and 6, we analyze all possible two and three-point correlator types and also comment on the kind of defect operators that can appear in a bulk-to-defect expansion of a given bulk operator. Having constructed all possible invariant structures,  in the next two sections we discuss $n$-point correlators involving $n_1$ bulk and $n_2$ defect operators and also parity-odd operators. To make the analysis complete we also add a brief discussion on how to get physical space components from the result in embedding space. In section 10, we extend the formalism to defects transforming in arbitrary representations of $SO(q)$. We finally end with a discussion of our results and outlook in section 11.
\paragraph{Note:} During the final stages of this manuscript, \cite{ope2} appeared on the arxiv which has an overlap with our work for the case of non-singlet defects.
\section{Formalism}
\subsection{Encoding Tensors as Polynomials}
We present a very quick review of the process of encoding tensors as polynomials in this section. For a detailed analysis the reader may refer to \cite{mixed1}. The encoding of tensors as polynomials makes computation much easier to handle. Consider a generic mixed symmetry representation of the $SO(d+1,1)$ group given by a Young diagram:
\begin{center}
$\lambda=$
\ytableausetup
{mathmode, boxsize=1.5em}
\begin{ytableau}
{} & {} & {} & \none[\dots]
& \scriptstyle {} & {} \\
{} & {} & {} & \none[\dots]
& {} \\
\none[\vdots] & \none[\vdots]
& \none[\vdots] \\
\scriptstyle {} & {} \\
{}
\end{ytableau} .
\end{center}
The Young diagram can be parametrized in two ways. The first way is to provide the heights of columns $h\equiv(h^{(1)}, h^{(2)}, \dots, h^{(n^C)})$, where $h^{(i)}$ is the height of the $i^{th}$ column and $n^C$ is the total number of columns. The second way is to provide the lengths of rows $l\equiv(l^{(1)}, l^{(2)}, \dots, l^{(n^R)})$, where $l^{(i)}$ is the length of the $i^{th}$ row and $n^R$ is the total number of rows. Given these parametrizations, the total number of boxes is given by,
\begin{equation}
    |\lambda|=\sum_{i=1}^{n_C} h^{(i)}=\sum_{i=1}^{n_R} l^{(i)} .
\end{equation}
A mixed symmetric tensor can be encoded as a polynomial by contracting its indices using one of the two sets of auxiliary vectors $\boldsymbol{\theta}=(\theta^{(1)}, \theta^{(2)},\dots, \theta^{(n^C)})$ and $\boldsymbol{z}=(z^{(1)}, z^{(2)}, \dots, z^{(n^R)})$. The vectors $\boldsymbol{\theta}$ are anti-commuting and encode the polynomial in an anti-symmetric basis, while the vectors $\boldsymbol{z}$ are commuting and encode the polynomial in a symmetric basis. Across a row $\boldsymbol{z}$ vector remains the same and down a column $\boldsymbol{\theta}$ remains the same. As an example for both bases,
\begin{center}
\ytableausetup{centertableaux}
\begin{ytableau}
z_1 & z_1 & z_1 \\
z_2 & z_2 & z_2\\
\end{ytableau}
\qquad 
\ytableausetup{centertableaux}
\begin{ytableau}
\theta_1 & \theta_2 & \theta_3\\
\theta_1 & \theta_2 & \theta_3\\
\end{ytableau} .
\end{center} 
A given Young representation is symmetric along the rows and anti-symmetric along the columns. Separate columns (rows) are symmetric (anti-symmetric) among themselves. The grassmanian nature of  $\boldsymbol{\theta}$-vectors is the following,   
\begin{equation}
    \theta^{(i)}_{m}\theta^{(j)}_{n}=(-1)^{\delta_{ij}}\theta^{(j)}_{n}\theta^{(i)}_{m} ,
\end{equation}
where indices $m$ and $n$ label the components of the auxiliary vectors. This relation encodes the anti-symmetry of $\boldsymbol{\theta}$-vectors only within the same column. We choose to do anti-symmetrization first using $\boldsymbol{\theta}$-vectors and then impose symmetrization by the action of $(z \cdot \partial_\theta)$ derivatives. Therefore, a mixed symmetry tensor can be encoded as:
\begin{equation}\label{symderiv}
   \tilde{f}(\boldsymbol{z})=\prod_{i=1}^{n_R} \prod_{j=1}^{\text{min}(l^{(i)},n^C)} \left( z^{(i)}\cdot \partial_{\theta^{(j)}}\right)f(\boldsymbol{\theta}) ,
\end{equation}
where,
\begin{equation}
    f(\boldsymbol{\theta})=\theta^{(1)}_{m_1}\dots \theta^{(1)}_{m_{h_1}}\theta^{(2)}_{m_{h_1+1}}\dots\theta^{(2)}_{m_{h_1+h_2}}\dots\theta^{(n_C)}_{m_{h_1+...+h_{n_C-1}+1}}\dots\theta^{(n_C)}_{m_{|\lambda|}}f^{m_1...m_{|\lambda|}} .
\end{equation}
The tracelessness condition can be imposed by demanding that certain dot products vanish:
\begin{equation}
\begin{split}
    f^{m_1...m_{|\lambda|}}\hspace{0.1cm} \text{traceless}&\Longleftrightarrow f(\boldsymbol{\theta})|_{\theta^{(i)}.\theta^{(j)}=0}\\
    &\Longleftrightarrow \tilde{f}(\boldsymbol{z})|_{z^{(i)}.z^{(j)}=0}.
\end{split}    
\end{equation}
To explicitly see the procedure of encoding tensors as polynomials, we consider two examples involving a symmetric two-tensor $S^{(mn)}$ and an anti-symmetric two-form $B^{[mn]}$. The two representations are,
\begin{center}
$A^{(mn)}$ =
\ytableausetup{centertableaux,boxsize=2em}
\begin{ytableau}
\theta^{(1)} & \theta^{(2)}  \\
\end{ytableau}
\qquad 
$B^{[mn]}$ =
\ytableausetup{centertableaux,boxsize=2em}
\begin{ytableau}
\theta^{(1)} \\
\theta^{(1)} \\
\end{ytableau} .
\end{center} 
We first convert the tensors into polynomial by contracting them with appropriate $\boldsymbol{\theta}$-vectors,
\begin{equation}
    \begin{split}
        A^{mn} \rightarrow A(\boldsymbol{\theta})=A^{mn}\theta^{(1)}_m\theta^{(2)}_n, \qquad B^{mn} \rightarrow B(\boldsymbol{\theta})= B^{mn}\theta^{(1)}_m\theta^{(1)}_n .
    \end{split}
\end{equation}
Once the polynomials have been constructed in $\boldsymbol{\theta}$-basis, symmetrization can be applied ($\ref{symderiv}$),
\begin{equation}
    \begin{split}
        \left(z^{(1)}\cdot \partial_{\theta^{(1)}}\right)\left(z^{(1)}\cdot \partial_{\theta^{(2)}}\right)A(\boldsymbol{\theta}), \qquad \left(z^{(1)}\cdot \partial_{\theta^{(1)}}\right)\left(z^{(2)}\cdot \partial_{\theta^{(1)}}\right)B(\boldsymbol{\theta}) .
    \end{split}
\end{equation}
Evaluating them we obtain the following result,
\begin{equation}
    \begin{split}
        A^{mn}z^{(1)}_m z^{(1)}_n, \qquad B^{mn}(z^{(1)}_m z^{(2)}_n-z^{(1)}_n z^{(2)}_m) .
    \end{split}
\end{equation}
Both the symmetric and anti-symmetric properties of the tensors have been captured.

So far, we have encoded a mixed symmetric tensor in the $d$-dimensional physical space where the CFT lives. In the next section, we will encode the tensor in a higher dimensional space-time where the action of the conformal group becomes linear.
\subsection{Embedding Formalism}\label{embedform}
We will briefly review the embedding space formalism and the procedure to encode mixed symmetric operators as polynomials in this space. For a detailed description of embedding space formalism, we refer the reader to \cite{spinning,mixed1}. The conformal group of a $d$-dimensional Euclidean CFT is $SO(d+1,1)$. This is also the Lorentz group in a $(d+2)$-dimensional Minkowski space. The $(d+2)$-dimensional space-time which we refer to as embedding space is the natural space associated with conformal transformations \cite{dirac}. The non-linear action of a conformal transformation in $d$-dimensional space becomes a linear Lorentz transformation in the embedding space. Let $P$ denote the coordinates of the embedding space. Points in the physical space are identified with null rays in the embedding space,
\begin{equation}
    P^2=0, \hspace{1cm} P\sim \alpha P \quad \text{where $\alpha \in \mathbb{R}^+$} .
\end{equation}
The first relation implies that everything in the theory lives on the light cone. We adopt lightcone coordinates to represent points on the cone. The second relation implies a gauge freedom in the identification of $P$ up to re-scaling. We can fix this gauge by setting $P^{+}=1$. This slice of the null cone is known as the Poincar\'{e} section. Physical points in $x\in\mathbb{R}^d$ are mapped to null points in this  Poincar\'{e} section:
\begin{equation}
   x \rightarrow P^{M}|_{x}=(P^{+},P^{-},P^{m})=(1,x^2,x^m) .
\end{equation}
The metric of the embedding space is the Lorentzian metric of $(d+1,1)$ space-time,
\begin{equation}
    P\cdot P=\eta_{MN}P^MP^N=-P^+P^-+\delta_{mn}P^mP^n .
\end{equation}
Operators in the physical space can be lifted to the embedding space. Consider a mixed symmetry tensor $f_{m_1...m_{|\lambda|}}(x)$ of dimension $\Delta$ in the physical space. This tensor can be uplifted to $F_{M_1...M_{|\lambda|}}(P)$ in the embedding space and satisfies the following conditions:
\begin{itemize}
    \item Homogeneity: $F_{M_1...M_{|\lambda|}}(\alpha P)=\alpha^{-\Delta}F_{M_1...M_{|\lambda|}}(P)$,
    \item Transversality: $P^{M_i}F_{M_1...M_i...M_{|\lambda|}}=0$ .
\end{itemize}
Operators in embedding space can once again be encoded as polynomials. We will use the auxiliary vectors $\boldsymbol{\Theta}=(\Theta^{(1)}, \Theta^{(2)},\dots, \Theta^{(n^C)})$ to encode anti-symmetry and \\$\boldsymbol{Z}=(Z^{(1)}, Z^{(2)}, \dots, Z^{(n^R)})$ to encode symmetry of the indices. We choose to write polynomials in the anti-symmetric basis (or $\boldsymbol{\Theta}$-basis) first and impose symmetrization via derivatives,
\begin{equation}\label{mixed101}
   \tilde{F}(P,\boldsymbol{Z})=\prod_{i=1}^{n_R} \prod_{j=1}^{\text{min}(l^{(i)},n^C)} \left( Z^{(i)}.\partial_{\Theta^{(j)}}\right)F(P,\boldsymbol{\Theta}),
\end{equation}
where,
\begin{equation}
    F(P,\boldsymbol{\Theta})=\Theta^{(1)}_{M_1}\dots \Theta^{(1)}_{M_{h_1}}\Theta^{(2)}_{M_{h_1+1}}\dots\Theta^{(2)}_{M_{h_1+h_2}}\dots\Theta^{(n^C)}_{M_{h_1+...+h_{n^C-1}+1}}\dots\Theta^{(n^C)}_{M_{|\lambda|}}F^{M_1...M_{|\lambda|}}(P).
\end{equation}
Once again the tracelessness condition can be encoded by demanding that certain dot products vanish:
\begin{equation}
\begin{split}
    F^{M_1\cdots M_{|\lambda |}}(P) \text{ \hspace{0.1cm}traceless/ transverse}&\Longleftrightarrow F(\boldsymbol{\Theta})|_{\Theta^{(p)}\cdot\Theta^{(q)}=0, P \cdot \Theta^{(p)}=0}\\
     &\Longleftrightarrow \tilde{F}(\boldsymbol{Z})|_{Z^{(p)}\cdot Z^{(q)}=0, P.Z^{(p)}=0} .
\end{split}    
\end{equation}
The $\Theta$ and $Z$ vectors satisfy the following properties,
\begin{equation}
    \Theta^{(i)}_a \cdot \Theta ^{(j)}_a=0, \quad Z^{(i)}_a \cdot \Theta^{(j)}_a=0, \quad Z^{(i)}_a \cdot Z^{(j)}_a=0 .
\end{equation}
The subscript refers to the operator the auxiliary vectors are associated with while the superscript on the auxiliary vectors indicates the column(row) for the $\boldsymbol{\Theta}(\boldsymbol{Z})$-vectors. In a given Young representation, one $\boldsymbol{Z}$-vector is used for contractions across a row while one $\boldsymbol{\Theta}$-vector is used for contractions along a column. The transversality condition also means that any polynomial constructed out of $\boldsymbol{\Theta}$ and $\boldsymbol{Z}$-vectors should be invariant under the following shift,
\begin{equation}
    \Theta_a^{(i)} \rightarrow \Theta_a^{(i)} + \alpha^{(i)} P_a, \quad Z^{(i)}_a \rightarrow Z^{(i)}_a + P_a .
\end{equation}
Here $\alpha^{(i)}$ carries the same Grassmanian signature as $\Theta^{(i)}$. Any quantity constructed out of $\boldsymbol{\Theta}$ and $\boldsymbol{Z}$ must be invariant under this symmetry as well. In the rest of the paper we will construct invariant objects out of $\boldsymbol{\Theta}$ that satisfy the transversality and tracelessness condition. Transversality implies that the product of the auxiliary vectors with their respective $P$ also vanish:
\begin{equation}
    P_a \cdot \Theta^{(i)}_a = 0, \quad P_a \cdot Z^{(i)}_a=0 .
\end{equation}
It is convenient to build all the invariant structures using $C_{MN}$ (C-tensor) which is transverse by construction,
\begin{equation}
    C^{{(i)}MN}_a = P_a^M\Theta^{{(i)}N}-P_a^N\Theta^{{(i)}M} .
\end{equation}
$C^{{(i)}MN}$ is also the smallest unit of $\Theta^{(i)}$ that satisfies transversality. A similar C-tensor can be constructed out of $\boldsymbol{Z}$-vectors. All other invariant structures will be constructed by contractions of C-tensor with various position vectors ($P_a$) and C-tensors. Contractions of more than two $C^{{(i)}MN}$ can be written in terms of contractions of two $C^{{(i)}MN}$,
\begin{equation}
    C_1^{MP}C_{2PR}C_3^{RN}=-\frac{1}{2}(C_1^{PR}C_{2PR})C^{MN}_1 .
\end{equation}
Therefore, we do not need to go beyond two C-tensor terms. To recover the uncontracted notation of tensors from a polynomial, we act on them with the following Todorov differential operator,
\begin{equation}\label{todo}
    D_{M}=\frac{d-2}{2}\frac{\partial}{\partial \Theta^M}+\Theta \cdot \frac{\partial}{\partial \Theta} \frac{\partial}{\partial \Theta^M} .
\end{equation}
Todorov differential operator is constructed to recover traceless symmetric tensors from polynomials \cite{todorov}. To recover free-indices for a spin-$l$ operator we apply the derivative $l$ times
\begin{equation}
    O_{M_1\cdots M_l}(P)=\frac{1}{l!(d/2-1)_l}D_{M_1}\cdots D_{M_l}O^l(P,\Theta^1\cdots \Theta^l).
\end{equation}
Here $(a)_l$ is the Polchhammer symbol. As discussed earlier, while constructing polynomials we first use an anti-symmetric basis  and then apply derivatives to impose the symmetrizations. An equally valid approach would be to first write everything in a symmetric basis and then apply the anti-symmetrization via derivatives. We will commit to using the former approach for the rest of the paper. Owing to its inherent anti-symmetry, the $\boldsymbol{\Theta}$ basis usually has a lower number of tensor structures compared to $\boldsymbol{Z}$-basis.  We reiterate that after taking the derivatives and projecting the results back to $d$-dimensions, the final result is basis-independent. 

To encode conserved operators we need an additional constraint. Conserved operators in physical space satisfy,
\begin{equation}
    \partial_{m}S^{mn \cdots} = 0 .
\end{equation}
To implement this in embedding space, first we need to free an index from the polynomial expression. This is implemented by acting with the Todorov derivative operator ($\ref{todo}$). Once an index has been freed, it can be contracted with a regular partial derivative to impose the conservation. Schematically it looks like:
\begin{equation}
    \partial^M D_M S(\Theta) = 0 .
\end{equation}
A detailed discussion of conserved tensors with its subtleties is given in \cite{spinning}.

\section{Embedding Formalism with a Defect}
\subsection{Defect}
A defect is an extended object (operator) living in an ambient space. A $q$ co-dimension defect breaks the full $d$-dimensional conformal group $SO(d+1,1)$ into $SO(p+1,1)\times SO(q)$ where $p+q=d$. Following \cite{gadde,marco}, such a defect is naturally identified in the embedding space as a $q$-dimensional time-like hyperplane intersecting the null cone. Projecting the intersection onto the Poincar\'{e} section results in defect in the physical space. Orientation of a hyperplane in the embedding space can be specified by providing a set of $q$ vectors ($P_{\alpha},\: \alpha=1,\dots,q$) that are orthogonal to it. The vectors $P_{\alpha}$ satisfy the following properties,
\begin{equation}
    P_{\alpha}\cdot X=0, \qquad X\cdot X=0, \qquad P_{\alpha}\cdot P_{\beta}=\delta_{\alpha\beta},
\end{equation}
where X is a point on the null cone. The inner product between two vectors $X$ and $Y$ in the embedding space naturally splits into two separate inner products of the $SO(p+1,1)$ and $SO(q)$ group: 
\begin{equation} \label{dotsplit}
\begin{split}
    X\cdot Y&= \left(\eta_{MN}-P_{\alpha M}P_{\alpha N}\right)X^{M}Y^{N}+P_{\alpha M}P_{\alpha N}X^{M}Y^{N}.\\
\end{split}
\end{equation}
It is convenient to split the coordinates into two sets: the first $p+2$ coordinates that are parallel to the defect and the last $q$ coordinates that are transverse to the defect. We will use letters $A,B,...$ to label the former and $I,J,...$ to label the latter.
\begin{equation}
    M=(A,I) \qquad A=1,2,\dots,p+2 \qquad I=1,2,\dots,q
\end{equation}
The inner product (\ref{dotsplit}) can be denoted as,
\begin{equation} \label{dotsplit1}
\begin{split}
    X\cdot Y&= \left(\eta_{MN}-P_{\alpha M}P_{\alpha N}\right)X^{M}Y^{N}+P_{\alpha M}P_{\alpha N}X^{M}Y^{N},\\
            &= X\bullet Y + X \circ Y,
\end{split}
\end{equation}
where we have defined
\begin{equation}
\begin{split}
    &X\bullet Y \equiv \left(\eta_{MN}-P_{\alpha M}P_{\alpha N}\right)X^{M}Y^{N}\\
    &X\circ Y \equiv  P_{\alpha M}P_{\alpha N}X^{M}Y^{N}.
\end{split}
\end{equation}
The above definitions allow us to make contact with the split representation used in \cite{marco} to study defects. In the physical space, $X\cdot Y\longrightarrow -(1/2)(x-y)^2$. Therefore equation (\ref{dotsplit1}) is merely stating that the square of the distance between two points is equal to sum of the squares of parallel and orthogonal distance to the flat defect. The perpendicular distance of a point ($X$) from a defect is given by,
\begin{equation}
    P_{\alpha M}P_{\alpha N}X^{M}X^{N}=(P_{\alpha}\cdot X)(P_{\alpha}\cdot X)=X\circ X \, .
\end{equation}
Formally we denote a $q$ co-dimension defect as $D^q(P_{\alpha})$.
Projecting the intersection of the hyperplane and the null cone onto the Poincar\'{e} section yields either a flat or a spherical defect depending on the orientation of the hyperplane. We will briefly discuss the two types below.
\subsubsection{Flat Defect}
\begin{figure}[ht]
\centering
\includegraphics[width=10cm, height=6cm]{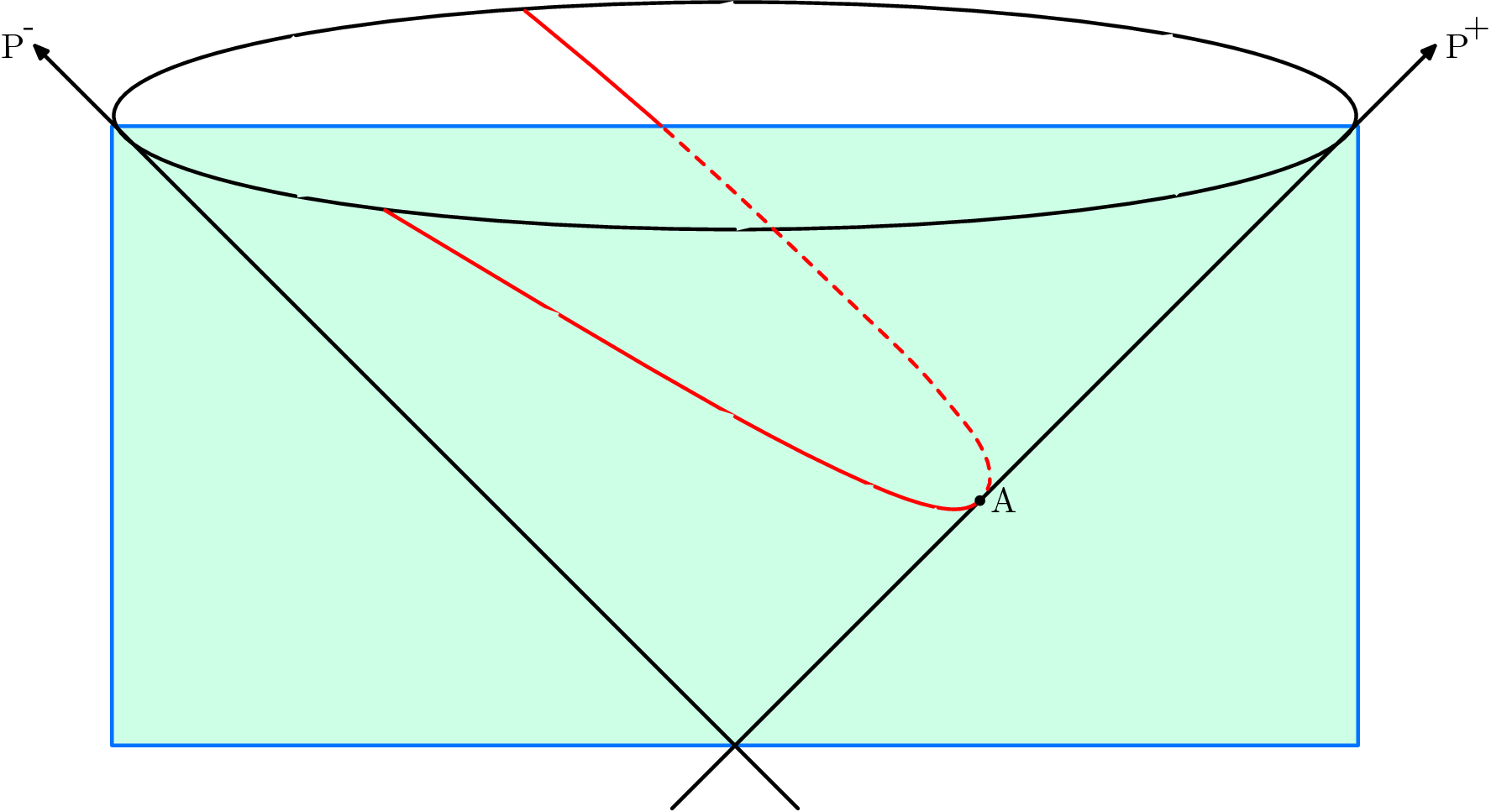}
\caption{The intersection of a defect hyperplane (containing the $P^+$-direction) with the Poincar\'{e} section (denoted here by point A) of the null-cone in embedding space leads to flat defect in the physical space.}
\end{figure}
A flat defect arises when the $P^+$-axis lies on the defect hyperplane. The intersection of the hyperplane with the Poincar\'{e} section results in only one point of intersection. Examples of flat defects include lines, planes and boundaries. Since $P^+$-axis lies on the hyperplane, we can conveniently choose the $P_{\alpha}$ vectors to be:
\begin{equation} \label{hyperplane}
P_{\alpha}=(\overbrace{0,\dots,0}^{p+2},\underbrace{0,...,1,...,0}_\text{$1$ at position $\alpha$}) \qquad \alpha=1,2,\dots,q.
\end{equation}
With the choice (\ref{hyperplane}) for $P_{\alpha}$ vectors, we get
\begin{equation}
\begin{split}
    &X\bullet Y \equiv \left(\eta_{MN}-P_{\alpha M}P_{\alpha N}\right)X^{M}Y^{N} = \eta_{AB}X^AY^B \\
    &X\circ Y \equiv  P_{\alpha M}P_{\alpha N}X^{M}Y^{N} = \delta_{IJ}X^IY^J.
\end{split}
\end{equation}
A bulk operator near a flat defect can be decomposed in terms of local operators living on the defect. This expansion is known as a bulk-to-defect expansion and in the embedding space looks like:
\begin{equation}\label{btd}
\Phi(P)|_D = \frac{b_{\Phi 1}}{(P\circ P)^{\Delta/2}}+\sum_{\hat{O}}\frac{b_{\Phi \hat{O}}\hat{O}(P)|_D}{(P\circ P)^{\frac{\Delta -\hat{\Delta}}{2}}} + \text{descendants} .
\end{equation}
Each defect local operator ($\hat{O}$) in ($\ref{btd}$) appears with a coupling-strength $b_{\Phi \hat{O}}$. This expansion is brought about by constructing a quantizing sphere centered on the defect and enclosing the bulk operator. The state on the sphere can then be shrunk to the center using scaling transformation resulting in defect local operators. Evaluating non-vanishing $\langle \Phi \hat{O} \rangle$ is essential for enumerating the representations that occur in this expansion.

\subsubsection{Spherical Defect}
\begin{figure}[ht]
\centering
\includegraphics[width=10cm, height=8.5cm]{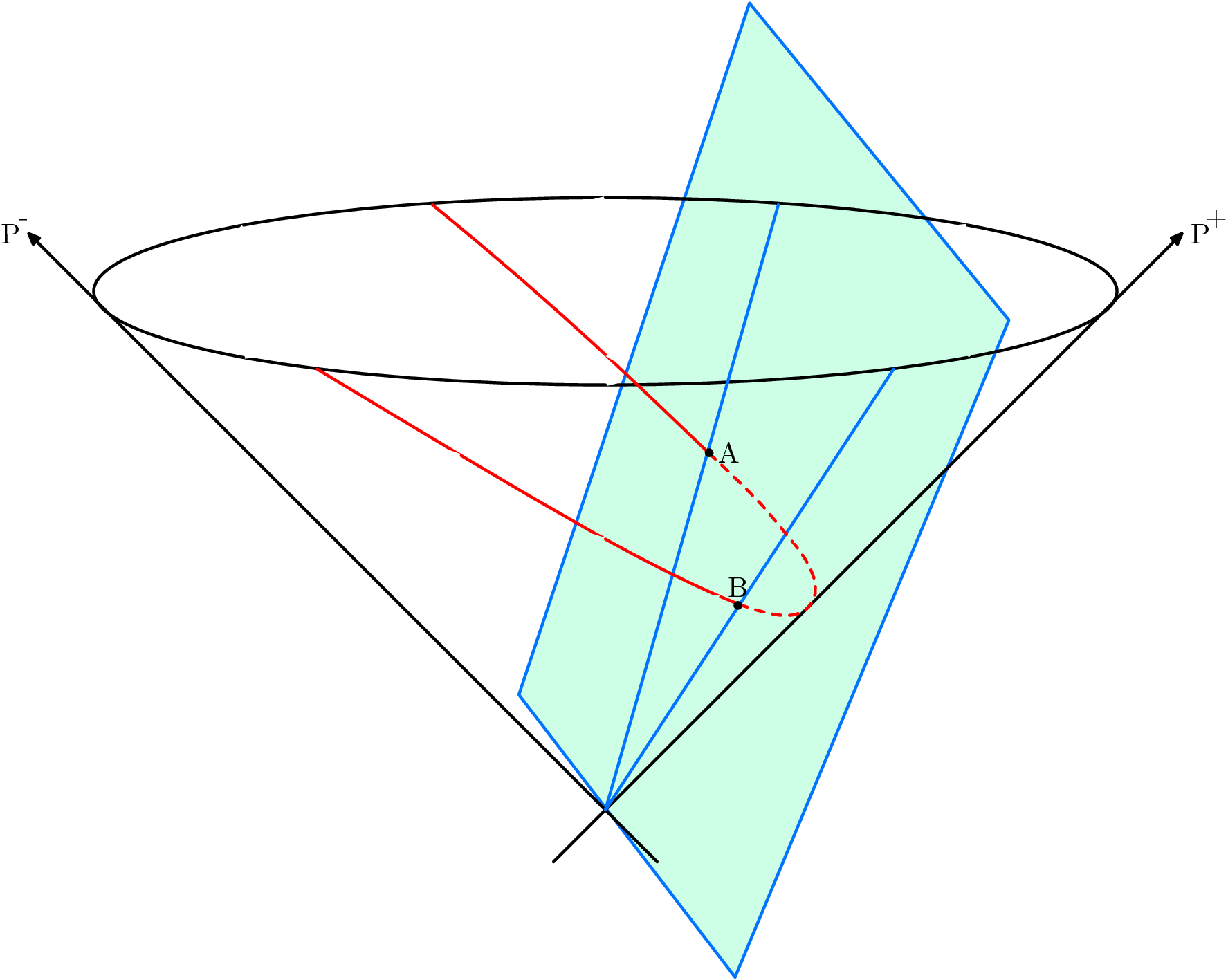} 
\caption{The intersection of a hyperplane (not containing $P^{+}-axis$) with the Poincar\'{e} section (denoted here by a pair of points A and B) of the null-cone in embedding space leads to spherical defect in the physical space.}
\label{fig:spherr}
\end{figure}
We obtain  a spherical defect when the defect hyperplane does not contain the $P^{+}$-axis. Spherical defects are characterized by their radius and center \footnote{
Since all lengths are relative in a conformal theory, the point at infinity (which is normalized as $\Omega=(0,1,0)$) as a reference point for the calculation of radius of the defect} \cite{gadde}. In addition to the bulk-to-defect expansion, there is an additional expansion channel known as the defect-to-bulk channel \cite{gadde}. A spherical defect can be written in terms of bulk primaries placed at the center of the defect. This channel is defined by enclosing the defect, and any operators on it, by a quantizing sphere. The projected state on this sphere can be shrunk down to a point (at the center of the defect) using scaling transformation. Schematically this is represented by,
\begin{equation}
D^q(P_{\alpha})=\sum_{\Phi}^{}c_{\Phi 1} \Phi(C) + \text{descendants} .
\end{equation}
In \cite{opedefect}, it was shown that in the limit where radius of the defect is very small ($R\rightarrow 0$),
\begin{equation}
D^q(P_{\alpha})=\sum_{\Phi}^{}c_{\Phi 1}R^{\Delta}\Phi(C) + O(R^2) .
\end{equation}
Following a similar procedure, it can be shown that including a defect local operator $\hat{O}$, sitting on the defect, in the defect-to-bulk expansion gives a similar result with a different coefficient,
\begin{equation}\label{defectwithlocal}
\hat{O}(Y)D^q(P_{\alpha}) =\sum_{\Phi}^{}c_{\Phi \hat{O}}R^{\Delta}\Phi(C) + O(R^2) .
\end{equation}
If multiple defect local operators are present, then the OPE of defect local operators can be used multiple times to reduce all of them in terms of a single defect operator.

\subsection{Formalism}
Having seen how to incorporate a defect in embedding space, let us now concentrate on defining operators and fields in presence of a defect. Throughout this work, our main focus will be on the flat defect case. However the results we present in this section are equally applicable to the case with spherical defects. The difference between the two defects arises when projecting the embedding space result back to physical space. We now have to deal with two kinds of operators: bulk operators and defect operators. Bulk operators transform under the complete group $SO(d+1,1)$ while the defect operators transform under the broken group $SO(d-q+1,1)\times SO(q)$. The uplift of a bulk operator to the (broken) embedding space will once again have to satisfy homogeneity, transversality and tracelessness conditions defined in the previous section. All the inner products split into two invariants (\ref{dotsplit1}). This implies,
\begin{equation}
    P_a\bullet\Theta_a^{(i)}=-P_a\circ\Theta_a^{(i)}, \hspace{1cm} \Theta_a^{(i)}\bullet\Theta_a^{(j)}=-\Theta_a^{(i)}\circ\Theta_a^{(j)} .
\end{equation}
Once again, we will use subscript in the embedding space vectors to identify different operators that might be under consideration. A similar relation holds for the $\boldsymbol{Z}$-vectors,
\begin{equation}
    Z^{(i)}_a \bullet Z^{(i)}_a=-Z^{(i)}_a\circ Z^{(i)}_a .
\end{equation}
Owing to the grassmanian nature of the $\boldsymbol{\Theta}$-vectors, in the split representation both of the products of $\Theta$-vectors vanish individually for $(i=j)$:
\begin{equation}\label{thetaprop}
    \Theta^{(i)}_a \bullet \Theta^{(i)}_a =0, \qquad \Theta^{(i)}_a \circ \Theta^{(i)}_a =0 .
\end{equation}
Since all the operators are on the null cone, 
\begin{equation}
    P_{a}\bullet P_{a}=-P_{a}\circ P_{a}.
\end{equation}
The C-tensor $C^{MN}$ introduced in the previous section breaks into three units \cite{marco} - $C^{AB}$, $C^{AI}$ and $C^{IJ}$. Fortunately not all of them are independent and these units follow the relation:
\begin{equation}
    \begin{split}
        &C_{AB}^{(i)}Q^AR^B=\frac{P\bullet R}{P\circ G}C_{AI}^{(i)}Q^AG^I-\frac{P\bullet Q}{P\circ G}C_{AI}^{(i)}R^AG^I,\\
        &C_{IJ}^{(i)}Q^IR^J=\frac{P\circ Q}{P\bullet G}C_{AI}^{(i)}G^AR^I-\frac{P\circ R}{P\bullet G}C_{AI}^{(i)}G^AQ^I.
    \end{split}
\end{equation}
The above relations imply that all invariant-structures can be built out of just $C_{AI}^{(i)}$. To make a polynomial in embedding space, we contract its indices with $\boldsymbol{\Theta}$-vectors. 

Defect local operators transform under the broken group $SO(p+1,1)\times SO(q)$. This implies that they carry separate quantum numbers corresponding to the parallel conformal group $SO(p+1,1)$ and the orthogonal group $SO(q)$. We will use auxiliary vectors $\{\Theta^{(i)}_{\hat{a}},Z^{(i)}_{\hat{a}}\}$ and $\{\Phi^{(j)}_{\hat{a}},W^{(i)}_{\hat{a}}\}$ corresponding to each broken group respectively. Position and auxiliary vectors associated with a defect operator are represented with a \textit{hat} symbol(e.g. $P_{\hat{a}}$). Enumerating all possible bulk and defect position and auxiliary vectors:
\begin{equation}
    P_a,P_{\hat{a}},\Theta^{(i)}_a,\Theta^{(i)}_{\hat{a}},\Phi^{(i)}_{\hat{a}},Z^{(i)}_{\hat{a}},W^{(i)}_{\hat{a}} .
\end{equation}
Since a defect local operator lies on the defect hyperplane, the vectors associated to it have the following properties:
\begin{equation}
    P_{\hat{a}I}=0, \quad \Theta^{(i)}_{\hat{a}I} = 0, \quad \Phi^{(i)}_{\hat{a}A}=0, \quad Z^{(i)}_{{\hat{a}}I}=0, \quad W^{(i)}_{{\hat{a}}A}=0 .
\end{equation}
The independent C-tensors for defect local operators are $C^{AB}$ and $C^{AI}$.
\begin{equation}
    \begin{split}
        &C_{\hat{a}}^{(i)AB}=P_{\hat{a}}^A\Theta_{\hat{a}}^{(i)B} - P_{\hat{a}}^B\Theta_{\hat{a}}^{(i)A} \\
        &C_{\hat{a}}^{(i)AI}=P_{\hat{a}}^A\Phi_{\hat{a}}^{(i)B}
    \end{split}    
\end{equation}
There would be C-tensor for the $\boldsymbol{Z}$-basis as well, however they only amount to replacing the $\boldsymbol{\Theta}$-vectors with $\boldsymbol{Z}$-vectors. In this work, we call transverse objects constructed out of C-tensors as \textit{invariants}. \textit{Invariants} will serve as building blocks for \textit{tensor structures}, which are the final structures appearing in correlators.

The number of independent invariants in $\boldsymbol{\Theta}$-basis can be obtained by considering all possible contractions between the position and auxiliary vectors that are under consideration (e.g. $P_a \bullet \Theta^{(i)}_{\hat{b}}$) minus the constraints imposed by demanding transversality of the auxiliary vector. Demanding transversality imposes a constraint for each $\Theta$-vector (bulk or defect) however $\Phi$-vectors impose no constraint as they are transverse by construction. If we ignore the fact that each auxiliary vector also has an $i$ index (labelling the column number for $\Theta$ vector), then given $n_1$ bulk operators and $n_2$ defect operators, the number of independent invariants is: 
\begin{equation}\label{numberofstructures} 
    3n_1^2-2n_1+2n_2^2-3n_2+5n_1n_2.
\end{equation}
We will provide another more rigorous derivation of the above relation in section ($\ref{secn}$) by listing down all possible independent invariants. It is essential to keep in mind that this relation only gives the number of independent invariants in the $\boldsymbol{\Theta}$-basis. The action of derivatives (to impose symmetrization) will reduce this number. 

Unless otherwise stated we will work with parity-even invariants and tensor structures.
Finally, we introduce a compact notation for position contractions involving bulk-bulk, bulk-defect and defect-defect operators.
\begin{equation}
    P_{ab}=(P_a \circ P_b) \quad P_{a\hat{b}}=(-2P_a\bullet P_{\hat{b}}) \quad P_{\hat{a}\hat{b}}=(-2P_{\hat{a}}\bullet P_{\hat{b}}) .
\end{equation}
An additional benefit of using the $\boldsymbol{\Theta}$-basis is that the dependence on the co-dimension of the defect is made manifest due to anti-symmetry.  The maximum number of a given $\Theta^{(i)}$ that can appear in a tensor structure is limited by the co-dimension of the defect ($q$). This limits the height of the Young representation.

\section{One-Point Correlators}
A distinguishing feature of a defect CFT is the non-vanishing nature of one-point correlators involving bulk local operators. Any bulk operator (near the defect) can be expanded in terms of defect operators ($\ref{defect expansion}$,$\ref{btd}$). Since a one-point correlator of identity operator is non-zero in a CFT, ($\ref{btd}$) implies that a one-point correlator of a bulk operator is non-zero. Consider a bulk operator in an arbitrary representation $\lambda$.
\begin{equation}
    \langle O_{\Delta,\lambda}(P_1,\boldsymbol{\Theta}_1 )\rangle 
\end{equation}
Only one invariant can be constructed with a single bulk operator,
\begin{equation}\label{onepointbasis}
    H_{1}^{(i,j)}=\frac{C_1^{(i)AI}C^{(j)}_{{1AI}}}{(P_1\circ P_1)} \quad \text{where $i\neq j$} .
\end{equation}
The parenthesis in $(i,j)$ does not imply symmetrization. If the number of columns of the operator representation is $l$, then taking into account the $i$-index in the above equation we find that there are $l(l-1)/2$ possible invariants. The tensor structures appearing in the correlation function must be constructed out of $H_{1}^{(i,j)}$ and should satisfy the homogeneity and transversality constraints:
\begin{equation}
    \langle O_{\Delta,\lambda}(P_1,\beta_1^{(i)}\Theta_1^{(i)} )\rangle=(\beta_1^{(1)})^{h^{(1)}}\cdots(\beta_1^{(l_1)})^{h^{(l_1)}}\langle O_{\Delta,\lambda}(P_1,\boldsymbol{\Theta}_1 )\rangle,
\end{equation}
\begin{equation}
    \langle O_{\Delta,\lambda}(\alpha P_1,\boldsymbol{\Theta}_1 )\rangle = \alpha^{-\Delta_1}\langle O_{\Delta,\lambda}(P_1,\boldsymbol{\Theta}_1 )\rangle.
\end{equation}
The final form of a one-point correlator is obtained after taking appropriate derivatives (to impose symmetrization),
\begin{equation}
    \langle O_{\Delta,\lambda}(P_1,\boldsymbol{Z}_1 )\rangle = \big(Z_1^{\lambda_1}\cdot \partial_{\Theta_1}^{\lambda_1}\big) \frac{T_{B}(\boldsymbol{\Theta}_1)}{(P_1\circ P_1)^{\Delta/2}} .
\end{equation}
$T_B(\boldsymbol{\Theta}_1)$ is an appropriate tensor structure satisfying homogeneity and transversality. Let us consider some specific cases.
\subsection{Symmetric Traceless}
We begin by considering spin-$l$ fields. The Young diagram for them is given as, 
\begin{center}
    \ytableausetup{centertableaux, boxsize=2em}
\begin{ytableau}
\Theta_1^{(1)} & \Theta_1^{(2)} & \cdots & \Theta_1^{(l)}\\
\end{ytableau}
\end{center}
One-point correlator has to be constructed out of ($\ref{onepointbasis}$). For a spin-$l$ operator we obtain,

\begin{equation}\label{onepointsym}
    \langle O_{\Delta,l}(P_1,Z_1^{(1)}) \rangle = \left(Z_1^{(1)}\cdot \partial_{\Theta_1^{(1)}}\right)\cdots \left(Z_1^{(1)}\cdot \partial_{\Theta_1^{(l)}}\right) \frac{T_B(\Theta_1,\Theta_2,\cdots,\Theta_l)}{(P\bullet P)^{\Delta/2}} .
\end{equation}
$T_B$ is a function of $H_1^{i,j}$. As an example, $T_B$ for spin 2 field is,
\begin{equation}\label{spin2}
    T_{B}(\Theta_1, \Theta_2)=H_1^{(1,2)} .
\end{equation}
For an odd-spin operator, it is not possible to write down any function that has the right homogeneity. Owing to this fact, a one-point correlator of an odd-spin operator is zero. Upon the application of the $Z\partial_{\Theta}$-derivatives on ($\ref{spin2}$), we obtain the following result,
\begin{equation}
    \langle O_{\Delta,l}(P_1,Z_1) \rangle=\frac{(H_{Z_1 Z_1})^{l/2}}{(P\circ P)^{\Delta/2}} .
\end{equation}
$l$ is even in the above case and $H_{Z_1Z_1}$ is ($\ref{onepointbasis}$) with $\Theta_1$ replaced by $Z_1$. When $q=1$ (boundary defect) we observe that $H_{Z_1 Z_1}=0$ and only the scalar operator has a non-zero one point correlator. This has been pointed out in multiple references (e.g. \cite{boundaryboots}). The $\boldsymbol{Z}$-derivatives are particularly simple for the traceless symmetric case and they only amount to replacing all the $\boldsymbol{\Theta}$-vectors with a single $\boldsymbol{Z}$-vectors. We will utilize this trick in all the symmetric traceless cases that we encounter.
\subsection{Forms}
We find that the one-point correlator of any $m$-form vanishes when considering parity-even invariants. However, this is not true if we consider parity-odd invariants. We will discuss parity-odd cases later in section ($\ref{podd}$). With just one $\Theta$, it is impossible to construct an invariant for a $m$-form.
\subsection{Two Column Operator}
Finally, we consider mixed symmetric operators in a two-column representation.
\begin{center}
    \ytableausetup{centertableaux, boxsize=2em}
\begin{ytableau}
\Theta_1^{(1)} & \Theta_1^{(2)} \\
\vdots & \vdots \\
\Theta_1^{(1)} & \Theta_1^{(2)} \\
\end{ytableau}
\end{center}
Both the columns have to be of equal height to obtain a non-zero correlator. For a two-column operator of height $h^{(1)}_1=h^{(2)}_1=h$ we obtain the following tensor structure,

\begin{equation}
    \frac{(H^{12}_1)^{h}}{(P\circ P)^{\Delta/2}} .
\end{equation}
The symmetrization will be imposed by the following derivative, 
\begin{equation}
     \left(Z_1^{(1)}\cdot \partial_{\Theta_1^{(1)}}\right)\left(Z_1^{(1)}\cdot\partial_{\Theta_1^{(2)}}\right)\cdots \left(Z_1^{(h)}\cdot \partial_{\Theta_1^{(1)}}\right)\left(Z_1^{(h)}\cdot\partial_{\Theta_1^{(2)}}\right) .
\end{equation}
An operator with $h=2$ (window operator) gives the following result after the action of symmetrization,
\begin{equation}
    H_{Z_1Z_1}H_{Z_2Z_2} - H_{Z_1Z_2}H_{Z_1Z_2} .
\end{equation}
When $q=2$, the above expression evaluates to zero. In general, for a $q$ co-dimension defect we get non-zero vacuum expectation value to a mixed symmetry operator of maximum height min$(q-1,d-q+1)$ \footnote{We thank Marco Meineri \cite{Lauria:2018klo} for pointing out this.}. In \cite{opedefect}, a duality between defects of different co-dimensions was pointed out:
\begin{equation}\label{duality}
    q \quad \Longleftrightarrow \quad d+2-q .
\end{equation}
We perform a check of this duality in terms of the height of an operator that can get a non-zero correlator.
\begin{center}
\begin{tabular}{ |c|c|c|c| } 
\hline
Dimension & Codimension & Height $h$ \\
\hline
$d+2-q$ & $q$ & min($q-1,d-q+1$) \\ 
$q$ & $d+2-q$ & min($d-q+1,q-1$) \\ 
\hline
\end{tabular}
\end{center}
\section{Two-Point Correlators}
Two and three-point correlators capture all the data of a defect CFT. In a defect CFT cross-ratios start appearing at the two-point (bulk) correlator level. This is the reason why bootstrap methods can be applied at this level. In this section we will list down two-point correlators. 
\subsection{Bulk-Defect}
We will first consider two-point correlators involving a bulk operator and a defect operator. These correlators are important as they contain information about the bulk and defect couplings. The defect operators that can appear in the bulk-to-defect expansion of a bulk operator ($\ref{defect expansion}$) can be identified by considering all non-zero bulk-defect two-point correlators. In fact, all possible operators appearing in the defect channel expansion can be found using the procedure given here. Consider the two point correlator, 
\begin{equation}
\langle O_{\Delta_1,\lambda_1} (P_1,\boldsymbol{\Theta}_1) \hat{O}_{\hat{\Delta},\lambda_2,\hat{\lambda}_2} (P_2,\boldsymbol{\Theta}_2,\boldsymbol{\Phi}_2)\rangle .
\end{equation}
The defect local operator has two representations corresponding to the two groups (parallel $\lambda$ and transverse $\bar{\lambda}$). For the defect operator, its vectors obey the following constraints:
\begin{equation*}
P_{\hat{2}}^I=0 \quad \Theta_{\hat{2}}^{(i)I} = 0\quad \Phi_{\hat{2}}^{(i)A}=0 .
\end{equation*}
We obtain the number of invariants to be 5 from ($\ref{numberofstructures}$). One of them was already present at one-point correlator level,
\begin{equation}
      H_a^{(i,j)} \quad \text{i $\neq$ j} .
\end{equation}
We encounter 4 additional invariants, 
\begin{equation} \label{invariantsBD}
\begin{split}
        &H^{(i,j)}_{a\hat{a}}=\frac{C_a^{(i)AB}C^{(j)}_{\hat{a}AB}}{P_a\bullet P_{\hat{a}}}, \qquad
        \: \: \: G_{a\hat{a}}^{(i)}=\frac{P_a\circ \Phi_{\hat{a}}^{(i)}}{(P_a\circ P_a)^{1/2}},\\
        &\tilde{G}_{a\hat{a}}^{(i,j)}=\frac{C_a^{(i)AI}P_{aA}\Phi_{\hat{a}I}^{(j)}}{(P_a\circ P_a)}, \qquad
        K_{a\hat{a}}^{(i)}=\frac{C_a^{(i)AI}P_{\hat{a}A}P_{aI}}{(P_a\circ P_a)^{1/2}(P_a\bullet P_{\hat{a}})} .
\end{split}
\end{equation}
Putting everything together we get the following invariants:
\begin{equation}
    H_{1\hat{2}}^{(i,j)} \quad H_1^{(i,j)} \quad G_{1\hat{2}}^{(i)} \quad \tilde{G}_{1\hat{2}}^{(i,j)}\quad K_{1\hat{2}}^{(i)} .
\end{equation}
With one bulk and one defect operator it is impossible to construct a cross-ratio. The final form of the correlator is,
\begin{equation}
    (Z_1^{\lambda_1}\cdot \partial_{\Theta_1}^{\lambda_1})(Z_2^{\lambda_{\hat{2}}}\cdot \partial_{\Theta_2}^{\lambda_{\hat{2}}})(W_2^{\bar{\lambda}_{\hat{3}}}\cdot \partial_{\Phi_2}^{\bar{\lambda}_{\hat{3}}}) \frac{T^{a}_{BD}b_a}{(-2P_1\bullet P_2)^{\hat{\Delta}}(P_1\circ P_1)^{(\Delta-\hat{\Delta})/2}} ,
\end{equation}
where,
\begin{equation}
   T^{a}_{BD} = \big(H_{1\hat{2}}^{(i,j)}\big)^{a_{ij}}\big(H_1^{i,j}\big)^{b_{ij}}\big(G_{1\hat{2}}^{i}\big)^{c_i}\big(\tilde{G}_{1\hat{2}}^{(i,j)}\big)^{d_{ij}}\big(K_{1\hat{2}}^{i}\big)^{e_i} .
\end{equation}
$T^{a}_{BD}$ refers to tensor structures and $b_a$ are the coefficients (bulk-to-defect) associated with each tensor structure. The derivatives are present to impose symmetrization on the tensor structures. Each invariant has a power associated with it, 
\begin{equation}
\begin{split}
    & H_{1\hat{2}}^{(i,j)}\rightarrow a_{ij}, \quad H_1^{i,j}\rightarrow b_{ij}, \quad G_{1\hat{2}}^{(i)}\rightarrow c_i, \quad \tilde{G}_{1\hat{2}}^{(i,j)}\rightarrow d_{ij}, \quad K_{1\hat{2}}^{(i)} \rightarrow e_i .
\end{split}
\end{equation}
We will set up some quick notations,
\begin{equation}\label{OhatOsystem}
    \begin{split}
        &n^C_{1}= \:\text{number of columns of $O$}\qquad h_1^{(i)} = \: \text{length of i-th column of $O$} \\
        &n^C_{\hat{2}}= \:\text{number of columns of $\hat{O}$ (parallel)}\quad h_{\hat{2}}^{(i)} = \: \text{length of i-th column of $\hat{O}$ (parallel)} \\
        &\bar{n}^C_{\hat{2}}= \:\text{number of columns of $\hat{O}$ (transverse)}\quad \bar{h}_{\hat{2}}^{(i)} = \: \text{length of i-th column of $\hat{O}$ (transverse)} .\\
    \end{split}
\end{equation}
The powers are subject to following conditions :
\begin{equation}\label{diph1}
    \begin{split}
        &h_1^{(i)} = \sum_j^{n^C_{\hat{2}}} a_{ij}+\sum_j^{n^C_{1}}b_{ij}+\sum_{j}^{\bar{n}^C_{\hat{2}}}d_{ij}+e_i ,\\
        &h_{\hat{2}}^{(i)}=\sum_j^{n^C_{1}}a_{ji},\\
        &\bar{h}_{\hat{2}}^{(i)}=c_i+\sum_{j}^{n^C_{1}}d_{ji} .
    \end{split}
\end{equation}
These equations have been determined by matching homogeneity of the invariants with that of the operators in the correlator. The solution for each variable has to be a non-negative integer and can be worked out easily. \textit{Mathematica} has a Reduce command which solves for integer solutions. We list down the relevant systems of equations for other correlators in the appendix. The system of equations can have multiple solutions. Each solution corresponds to a different tensor structure which can appear with a different coefficient. Once the tensor structures have been computed, they need to be acted on by the appropriate symmetrization derivatives.

Let us consider a concrete example:
\begin{equation}
\langle O_{\Delta_1,\lambda_1} (P_1, \Theta_1) \hat{O}_{\hat{\Delta},\lambda_2,\hat{\lambda}_2}  (P_2,\Phi_2)\rangle .
\end{equation}
We consider a two-point correlator between a bulk vector and a defect operator with spin-1 orthogonal to the defect,
\begin{center}
$\lambda_1 =$
\ytableausetup{centertableaux}
\begin{ytableau}
  \\ 
\end{ytableau}
\hspace{1cm}$\lambda_{\hat{2}} =$
$\bullet$
\hspace{1cm} $\bar{\lambda}_{\hat{2}}=$
\ytableausetup{centertableaux}
\begin{ytableau}
  \\ 
\end{ytableau} .
\end{center}
Plugging $h^{(1)}_1=1$, $h^{(1)}_{\hat{2}}=0$ and $\bar{h}^{(1)}_{\hat{2}}=1$ in ($\ref{diph1}$) we obtain two tensor structures,
\begin{equation}
\langle O_{\lambda_1} (P_1,\Theta_1) \hat{O}_{\lambda_{\hat{2}},\bar{\lambda}_{\hat{2}}} (P_{\hat{2}},\Phi_{\hat{2}})\rangle=\frac{b_1\tilde{G}_{1\hat{2}}+b_2K_{1\hat{2}}G_{1\hat{2}}}{(-2P_1\bullet P_{\hat{2}})^{\hat{\Delta}}(P_1\circ P_1)^{(\Delta-\hat{\Delta})/2}}.
\end{equation}
We can further demand that the bulk operator is a conserved spin-$1$ current with dimension $\Delta_1=d-1$. Conservation condition implies,
\begin{equation}
    \partial^M D_M \langle O_{\lambda_1} (P_1,\Theta_1) \hat{O}_{\lambda_{\hat{2}},\bar{\lambda}_{\hat{2}}} (P_{\hat{2}},\Phi_{\hat{2}})\rangle=0.
\end{equation}
This results in a relation between the coefficients $b_1$ and $b_2$:
\begin{equation}
    b_1(q-1)+b_2(q-d+\hat{\Delta})=0.
\end{equation}
We will now list down the representations that can occur in the decomposition of different bulk operators.
\newline
\newline
\newline
\textbf{Scalar Bulk Operator}
\newline
We consider all possible two-point correlators with a bulk scalar. The correlator is non-zero in the following case only,
\begin{equation}
    \langle O_{\Delta}(P_1) \hat{O}_{\hat{\Delta},0,s}(P_2,W_2)\rangle .
\end{equation}
where $s$ is a symmetric traceless quantum number of the $SO(q)$ representation. This indicates that a bulk scalar decomposes into defect local operators transforming as symmetric traceless tensors under $SO(q)$  while being scalars under the  $SO(p+1,1)$ group. Schematically this can be represented as,
\begin{equation}\label{scalardecomp}
    O \sim \hat{O}+\hat{O}^i +\hat{O}^{(ij)}+\dots
\end{equation}
\textbf{Spin-$\ell$ Bulk Operator}
\newline
The defect decomposition of a spin-$\ell$ bulk operator yields defect operators in the following representations.
\begin{center}
\begin{tabular}{ |c|c|c|c| } 
\hline
Spin-$J$ & Spin-$j$ & Height of $\bar{\lambda}_{\hat{2}}$ \\
\hline
\multirow{3}{4em}{Spin-$\ell$} & Spin-$\ell$ & 1 \\ 
& Spin $\ell$-1 & $\ell$ \\ 
& \vdots & $\ell$ \\
& 0 & $\ell$ \\
\hline
\end{tabular}

\end{center}
Spin-$J$ represents the spin of the bulk operator, spin-$j$ represents the spin of the defect operator parallel to the defect and last column represents the maximum height of the $SO(q)$ representation of the defect operator. For a spin-$\ell$ bulk primary, we find that the defect operators appearing in the defect expansion are spinning fields under the $SO(p+1,1)$ group while the maximum height of the representation under $SO(q)$ is restricted by $\ell$. The height of the  $SO(q)$ representation is also limited by the co-dimension of the defect. It can have a maximum height of $q$ (irrespective of $\ell$). If the co-dimension of the defect is 1, then the only operators occurring would transform in the traceless symmetric representation of $SO(q)$. 
\subsection{Bulk-Bulk}\label{BBsec}
We will now consider bulk-bulk two-point correlators. The conformal symmetry does not completely fix the position dependence of the correlator. The following two conformal cross-ratios \cite{marco} can be constructed:
\begin{equation}
    \xi_1=\frac{2P_1 \bullet P_2}{(P_1\circ P_1)^{1/2}(P_2\circ P_2)^{1/2}} , \hspace{1cm} \xi_2= \frac{2P_1\circ P_2}{(P_1\circ P_1)^{1/2}(P_2\circ P_2)^{1/2}} .
\end{equation}
With these two cross-ratios, the bulk-bulk two-point correlator can be written as:
\begin{equation}
    \langle O_{\Delta_1,\lambda_1} (P_1,\boldsymbol{\Theta}_1) O_{\Delta_2,\lambda_2} (P_2,\boldsymbol{\Theta}_2)\rangle=\sum_n\frac{T_{BB}^{(n)}f_{n}(\xi_1,\xi_2)}{(P_1\boldsymbol{\circ}P_1)^{\Delta_1/2}(P_2\boldsymbol{\circ}P_2)^{\Delta_2/2}},
\end{equation}
where $T^{(n)}_{BB}$ are the different tensor structures compatible with the representation of the operators and the functions $f_n(\xi_1,\xi_2)$ can be expanded in terms of bulk-channel conformal blocks. In case of only bulk operators in a correlation function, the invariants that can appear are of the form:
\begin{equation} \label{invariantsBB}
\begin{split}
   &\qquad \qquad  \qquad \qquad \quad  H_{a}^{(i,j)}=\frac{C_a^{(i)AI}C_{aAI}^{(j)}}{(P_a\circ P_a)},\\
   &S_{ab}^{(i,j)}=\frac{C_a^{(i)AI}C_b^{(j)BI}P_{aA}P_{bB}}{(P_a\circ P_a)(P_b\circ P_b)}, \qquad \Bar{S}_{ab}^{(i,j)}=\frac{C_a^{(i)AI}C_b^{(j)AJ}P_{aI}P_{bJ}}{(P_a\circ P_a)(P_b\circ P_b)},\\
   &K_{ab}^{(i)}=\frac{C_a^{(i)AI}P_{aA}P_{bI}}{(P_a\circ P_a)(P_b\circ P_b)^{1/2}}, \qquad \Bar{K}_{ab}^{(i)}=\frac{C_a^{(i)AI}P_{bA}P_{bI}}{(P_a\circ P_a)^{1/2}(P_b\circ P_b)},
\end{split}
\end{equation}
with $a\neq b$ in all the above invariants. The above invariants have the following properties:
\begin{equation} \label{relationsBB}
    \begin{split}
        &H_a^{(i,j)}=H_a^{(j,i)},\\
        &S_{(ab)}^{(i,j)}=S_{(ba)}^{(j,i)},\\
        &\Bar{S}_{(ab)}^{(i,j)}=\Bar{S}_{(ba)}^{(j,i)}.
    \end{split}
\end{equation}
It is also possible to construct additional invariants like,
\begin{equation}
    \frac{C_a^{(i)AI}P_{bA}P_{cI}}{(P_a\circ P_a)^{1/2}(P_b\circ P_b)^{1/2}(P_c\circ P_c)^{1/2}} \quad \text{where ($a\neq b\neq c$) } .
\end{equation}
However, the above invariant can be shown to be a linear combination of the invariants already defined in (\ref{invariantsBB}) using identities listed in ($\ref{Bident}$).
The list of independent invariants is:
\begin{equation}
    H_1^{(i,j)}, H_2^{(i,j)}, S_{12}^{(i,j)}, \Bar{S}_{12}^{(i,j)}, K_{12}^{(i)}, K_{21}^{(i)}, \Bar{K}_{12}^{(i)}, \Bar{K}_{21}^{(i)} .
\end{equation}
Depending on the representation of bulk operators (including the $i$-index in the above equation), the total number of invariants is (we refer the reader to ($\ref{notations}$) for notations),
\begin{equation}
    \frac{1}{2}(l^{(1)}_{1}+l^{(1)}_{2})(l^{(1)}_{1}+l^{(1)}_{2}+3)+l^{(1)}_{1}l^{(1)}_{2} .
\end{equation}
The two-point correlator is non-zero only for identical operators in an ordinary CFT. This is no longer true in a defect CFT and two-point correlators between arbitrary operators can be non-zero. We will discuss two examples for the bulk-bulk two-point correlators. The system of equations to evaluate the tensor structures for a given two-point correlator is listed in the appendix ($\ref{OOappendix}$). We first consider a two-point correlator between a two-form and a vector.
\begin{center}
$\lambda_1 =$
\ytableausetup{centertableaux}
\begin{ytableau}
  \\ 
  \\
\end{ytableau}
\hspace{1cm}$\lambda_2 =$
\ytableausetup{centertableaux}
\begin{ytableau}
  \\ 
\end{ytableau}
\end{center}
Using the invariants (\ref{invariantsBB}), and applying the equations of ($\ref{OOappendix}$) with $h^{(1)}_1=2$ and $h^{(1)}_{\hat{2}}=1$ we get the following tensor structures:
\begin{equation} \label{exampleBB1}
    \begin{split}
        \sum_{n} T_{BB}^{(n)}f_{n}(\xi_1,\xi_2)=
        &K_{12}^{(1)}\Bar{K}_{12}^{(1)}K_{21}^{(1)}f_1(\xi_1,\xi_2)+K_{12}^{(1)}\bar{K}_{12}^{(1)}\Bar{K}_{21}^{(1)}f_2(\xi_1,\xi_2)+S_{12}^{(1,1)}K_{12}^{(1)}f_{3}(\xi_1,\xi_2)\\
        &+S_{12}^{(1,1)}\Bar{K}_{12}^{(1)}f_{4}(\xi_1,\xi_2)
        +\bar{S}_{12}^{(1,1)}K_{12}^{(1)}f_{5}(\xi_1,\xi_2)+\Bar{S}_{12}^{(1,1)}\Bar{K}_{12}^{(1)}f_{6}(\xi_1,\xi_2) .\\
    \end{split}
\end{equation}
The next step is to apply derivatives to complete the symmetrization,
\begin{equation}
    \begin{split}
        \left(Z_1^{(1)}.\partial_{\Theta_1^{(1)}}\right)\left(Z_1^{(2)}.\partial_{\Theta_1^{(1)}}\right)\sum_{n}T_{BB}^{(n)}f_{n}(\xi_1,\xi_2)&=\left(K_{12}^{(2_Z)}\Bar{K}_{12}^{(1_Z)}K_{21}^{(1)}-K_{12}^{(1_Z)}\Bar{K}_{12}^{(2_Z)}K_{21}^{(1)}\right)f_1(\xi_1,\xi_2)\\
        &+\left(K_{12}^{(2_Z)}\Bar{K}_{12}^{(1_Z)}\Bar{K}_{21}^{(1)}-K_{12}^{(1_Z)}\Bar{K}_{12}^{(2_Z)}\Bar{K}_{21}^{(1)}\right)f_2(\xi_1,\xi_2)\\
        &+\left(S_{12}^{(2_Z,1)}K_{12}^{(1_Z)}-S_{12}^{(1_Z,1)}K_{12}^{(2_Z)}\right)f_{3}(\xi_1,\xi_2)\\
        &+\left(S_{12}^{(2_Z,1)}\Bar{K}_{12}^{(1_Z)}-S_{12}^{(1_Z,1)}\Bar{K}_{12}^{(2_Z)}\right)f_{4}(\xi_1,\xi_2)\\
        &+\left(\Bar{S}_{12}^{(2_Z,1)}K_{12}^{(1_Z)}-\Bar{S}_{12}^{(1_Z,1)}K_{12}^{(2_Z)}\right)f_{5}(\xi_1,\xi_2)\\
        &+\left(\Bar{S}_{12}^{(2_Z,1)}\Bar{K}_{12}^{(1_Z)}-\Bar{S}_{12}^{(1_Z,1)}\Bar{K}_{12}^{(2_Z)}\right)f_{6}(\xi_1,\xi_2) .
    \end{split}
\end{equation}
The application of derivatives results in a lot of terms. Although this is correct, it is not required as the operators under consideration are not symmetric in their indices. When operators do not have symmetry (anti-symmetry), the result in $\boldsymbol{\Theta}$-basis ($\boldsymbol{Z}$-basis) is sufficient.

Let us look at another example involving a hook and a scalar operator.
\begin{center}
$\lambda_1 =$
\ytableausetup{centertableaux}
\begin{ytableau}
 {} & {} \\
 {} \\
\end{ytableau}
\hspace{1cm}$\lambda_2 =\bullet$
\end{center}
We again use ($\ref{OOappendix}$) with $h_1^{(1)}=2$, $h_1^{(2)}=1$ and $h^{(1)}_{\hat{2}}=0$ to find the tensor structures,
\begin{equation}
    \begin{split}
        \sum_{n} T_{BB}^{(n)}f_{n}(\xi_1,\xi_2)&=H_1^{(1,2)}K_{12}^{(1)}f_{1}(\xi_1,\xi_2)+H_1^{(1,2)}\Bar{K}_{12}^{(1)}f_{2}(\xi_1,\xi_2)\\
        &+K_{12}^{(1)}\Bar{K}_{12}^{(1)}K_{12}^{(2)}f_3(\xi_1,\xi_2)+K_{12}^{(1)}\Bar{K}_{12}^{(1)}\Bar{K}_{12}^{(2)}f_{4}(\xi_1,\xi_2) .
    \end{split}
\end{equation}
We apply the derivatives to symmetrize the tensor structures,
\begin{equation}
    \begin{split}
        \left(Z_1^{(2)}.\partial_{\Theta_1^{(1)}}\right)\left(Z_1^{(1)}.\partial_{\Theta_1^{(1)}}\right)&\left(Z_1^{(1)}.\partial_{\Theta_1^{(2)}}\right)\sum_{n}T_{BB}^{(n)}f_{n}(\xi_1,\xi_2)\\
        &=\left(H_1^{(1_Z,1_Z)}K_{12}^{(2_Z)}-H_1^{(2_Z,1_Z)}K_{12}^{(1_Z)}\right)f_{1}(\xi_1,\xi_2)\\
        &+\left(H_1^{(1_Z,1_Z)}\Bar{K}_{12}^{(2_Z)}-H_1^{(2_Z,1_Z)}\Bar{K}_{12}^{(1_Z)}\right)f_{2}(\xi_1,\xi_2)\\
        &+\left(K_{12}^{(1_Z)}\Bar{K}_{12}^{(2_Z)}K_{12}^{(1_Z)}-K_{12}^{(2_Z)}\Bar{K}_{12}^{(1_Z)}K_{12}^{(1_Z)}\right)f_3(\xi_1,\xi_2)\\
        &+\left(K_{12}^{(1_Z)}\Bar{K}_{12}^{(2_Z)}\Bar{K}_{12}^{(1_Z)}-K_{12}^{(2_Z)}\Bar{K}_{12}^{(1_Z)}\Bar{K}_{12}^{(1_Z)}\right)f_4(\xi_1,\xi_2) .\\
    \end{split}
\end{equation}
This is the final result for a two-point correlator involving a hook and a scalar operator. All the symmetries and anti-symmetries of the hook operator are made explicit after the action of derivative. 
\subsection{Defect-Defect}
We finally study two-point correlators of defect local operators,
\begin{equation}
    \langle \hat{O}_{\lambda_1,\hat{\lambda}_1} (P_{\hat{1}},\boldsymbol{\Theta}_1, \boldsymbol{\Phi_1}) \hat{O}_{\lambda_2,\hat{\lambda}_2} (P_{\hat{2}},\boldsymbol{\Theta}_2,\boldsymbol{\Phi_2})\rangle.
\end{equation}
We find new invariants constructed by the contraction of $\Theta$ and $\Phi$ among themselves:
\begin{equation} \label{invariantsDD}
    \begin{split}
        &H_{\hat{a}\hat{b}}^{(i,j)}=\frac{C_{\hat{a}}^{(i)AB}C_{\hat{b}}^{(j)AB}}{P_{\hat{a}}\bullet P_{\hat{b}}},\\
        &\Tilde{H}_{\hat{a}\hat{b}}^{(i,j)}=\Phi_{\hat{a}}^{(i)}\circ \Phi_{\hat{b}}^{(j)}.
    \end{split}
\end{equation}
Since we are considering defect operators, the condition $\hat{a} \neq \hat{b}$ is automatically implied. The possible invariants are  $H_{\hat{1}\hat{2}}^{(i,j)}$ and $\Tilde{H}_{\hat{1}\hat{2}}^{(i,j)}$ and this implies that the defect operators should have the same representation for both sectors.
\begin{equation}
    \lambda_{\hat{1}}=\lambda_{\hat{2}}, \hspace{1cm} \bar{\lambda}_{\hat{1}}=\bar{\lambda}_{\hat{2}}.
\end{equation}
This has to be true since defect local operators behave like operators of an ordinary CFT. There is no cross-ratio in this case as the conformal symmetry completely fixes the form of the correlator.
\section{Three-Point Correlators}
Crossing equations involving three-point correlators constrain the data-set of a defect CFT. These are analogous to four-point crossings in an ordinary CFT.
\subsection{Bulk-Bulk-Bulk}
No additional invariants appear for bulk three-point correlators and the ones listed in (\ref{invariantsBB}) are sufficient. The total number of invariants in this case is 21 from ($\ref{numberofstructures}$) and we list them below:
\begin{equation}
\begin{split}
    &H_1^{(i,j)}, H_2^{(i,j)}, H_3^{(i,j)},\\
    &S_{12}^{(i,j)}, S_{23}^{(i,j)}, S_{31}^{(i,j)}, \\
    &\Bar{S}_{12}^{(i,j)}, \Bar{S}_{23}^{(i,j)}, \Bar{S}_{31}^{(i,j)}, \\
    &K_{12}^{(i)}, K_{21}^{(i)}, K_{23}^{(i)}, K_{32}^{(i)}, K_{31}^{(i)}, K_{13}^{(i)},\\ &\Bar{K}_{12}^{(i)}, \Bar{K}_{21}^{(i)}, \Bar{K}_{23}^{(i)}, \Bar{K}_{32}^{(i)}, \Bar{K}_{31}^{(i)}, \Bar{K}_{13}^{(i)} .
\end{split}
\end{equation}
Depending on the representation of the bulk operators we determine the number of invariants (taking into account $i$-index) to be:
\begin{equation}
    \frac{1}{2}(l^{(1)}_{1}+l^{(1)}_{2}+l^{(1)}_{3})(l^{(1)}_{1}+l^{(1)}_{2}+l^{(1)}_{3}+7)+l^{(1)}_{1}l^{(1)}_{2}+l^{(1)}_{2}l^{(1)}_{3}+l^{(1)}_{3}l^{(1)}_{1} .
\end{equation}
Consider a three-point correlator,
\begin{equation}
    \langle O_{\lambda_1} (P_1,\boldsymbol{\Theta}_1) O_{\lambda_2} (P_2,\boldsymbol{\Theta}_2)O_{\lambda_3} (P_3,\boldsymbol{\Theta}_3)\rangle=\sum_n\frac{f^{(n)}_{123}T_{BBB}^{(n)}f_{n}(\xi_1,...,\xi_6)}{(P_{11})^{\Delta_1/2}(P_{22})^{\Delta_2/2}(P_{33})^{\Delta_3/2}} .
\end{equation}
Here $T_{BBB}^{(n)}$ are three-point tensor structures and functions $f_{n}(\xi_1,...,\xi_6)$ can be expanded in terms of three-point conformal blocks. The conformal blocks are functions of the cross-ratios. Six cross-ratios can be constructed out of three bulk operators:
\begin{equation}
    \begin{split}
        &\xi_1=\frac{2P_1 \bullet P_2}{(P_{11})^{1/2}(P_{22})^{1/2}} , \hspace{1cm} \xi_2= \frac{2P_1\circ P_2}{(P_{11})^{1/2}(P_{22})^{1/2}}\\
        &\xi_3=\frac{2P_2 \bullet P_3}{(P_{22})^{1/2}(P_{33})^{1/2}} , \hspace{1cm} \xi_4= \frac{2P_2\circ P_3}{(P_{22})^{1/2}(P_{33})^{1/2}}\\
        &\xi_5=\frac{2P_3 \bullet P_1}{(P_{33})^{1/2}(P_{11})^{1/2}} , \hspace{1cm} \xi_6= \frac{2P_3\circ P_1}{(P_{33})^{1/2}(P_{11})^{1/2}} .\\
    \end{split}
\end{equation}
As an example, let us consider  three-point correlator of a 2-form and two scalars.
\begin{center}
$\lambda_1 =$
\ytableausetup{centertableaux}
\begin{ytableau}
  \\ 
  \\
\end{ytableau}
\hspace{1cm}$\lambda_2 =$
$\bullet$
\hspace{1cm} $\lambda_3=$
\ytableausetup{centertableaux}
$\bullet$
\end{center}
We use the system of equations (obtained from homogeneity constraints) in ($\ref{OOOappendix}$) with $h^{(1)}_1=2$ and $h^{(1)}_{2}=h^{(1)}_3=0$ to obtain the tensor structures,
\begin{equation}
    \begin{split}
        \sum_n f^{(n)}_{123} T^{(n)}_{BBB}f_n(\xi)=&f^{(1)}_{123}K_{12}K_{13}f_1(\xi) + f^{(2)}_{123}K_{12}\bar{K}_{12}f_2(\xi) + f^{(3)}_{123}K_{12}\bar{K}_{13}f_3(\xi) \\ &+ f^{(4)}_{123}K_{13}\bar{K}_{12}f_4(\xi) + f^{(5)}_{123}K_{13}\bar{K}_{13}f_5(\xi) + f^{(6)}_{123}\bar{K}_{12}\bar{K}_{13}f_6(\xi) .
    \end{split}
\end{equation}
$f^{(n)}_{123}$ are three-point coefficients associated to each tensor structure. 
\subsection{Bulk-Bulk-Defect}
Three-point correlators involving two bulk and one defect operators are important for bootstrap as pointed out in \cite{marco2}. We encounter one new invariant in this case,
\begin{equation}
\begin{split}
   &N_{\hat{c},1b}^{(i)}=\frac{C_{\hat{c}}^{(i)AB}P_{1A}P_{bB}}{(P_{\hat{c}}\bullet P_1)^{1/2}(P_{1}\bullet P_b)^{1/2}(P_{b}\bullet P_{\hat{c}})^{1/2}} \quad (1 \neq b) . \\
\end{split}
\end{equation}
We determine the number of invariants to be 17 by plugging in ($n_1=2$ and $n_2=1$) in ($\ref{numberofstructures}$). We list them below,

\begin{equation}
\begin{split}
    &G_{1\hat{3}}^{(i)} , G_{2\hat{3}}^{(i)}, \\
    &H_1^{(i,j)},H_2^{(i,j)}, \\
    &\Bar{K}_{12}^{(i)},K_{12}^{(i)},\Bar{K}_{21}^{(i)},K_{21}^{(i)}, \\
    &\tilde{G}_{1\hat{3}}^{(i,j)},\tilde{G}_{2\hat{3}}^{(i,j)},H_{1\hat{3}}^{(i,j)},H_{2\hat{3}}^{(i,j)}, \\
    &S_{12}^{(i,j)},\Bar{S}_{12}^{(i,j)}, \\
    &K_{1\hat{3}}^{(i)},K_{2\hat{3}}^{(i)}, \\
    &N_{\hat{3},12}^{(i)} .
\end{split}
\end{equation}
There are three independent cross-ratios in this case. The two bulk operators yield two cross-ratios which we already encountered before  $\xi_1$ and $\xi_2$. Including the defect operator yields an additional cross-ratio,
\begin{equation}
    \chi=\frac{(P_{\hat{3}}\bullet P_1) (P_2 \bullet P_2)^{1/2}}{(P_{\hat{3}}\bullet P_2)(P_1 \bullet P_1)^{1/2}} .
\end{equation}
A three-point correlator involving two bulk and one defect operator has the following structure:
\begin{equation}
    \langle O_1 O_2 \hat{O}_3 \rangle = \sum_n \frac{T^{(n)}_{BBD}f_n(\xi_1,\xi_2,\chi)}{(P_{12})^{\frac{\Delta_1+\Delta_2-\Delta_3}{2}}(P_{1\hat{3}})^{\frac{\Delta_1+\hat{\Delta}_3-\Delta_2}{2}}(P_{2\hat{3}})^{\frac{\Delta_2+\hat{\Delta}_3-\Delta_1}{2}}} .
\end{equation}
As an example, let us consider a three-point correlator involving a vector, a scalar and a defect operator which is a 2-form along the defect and a scalar orthogonal to the defect.
\begin{center}
$\lambda_1 =$
\ytableausetup{centertableaux}
\begin{ytableau}
  \\ 
\end{ytableau}
\hspace{1cm}$\lambda_2 =$
$\bullet$
\hspace{1cm} $\lambda_{\hat{3}}=$
\ytableausetup{centertableaux}
\begin{ytableau}
  \\ 
  \\
\end{ytableau}
\quad \quad
$\bar{\lambda}_{\hat{3}}=\bullet$
\end{center}
Using the system of equations listed in ($\ref{OOhatOappendix}$) and taking $h^{(1)}_1=1$, $h^{(1)}_2=0$, $h^{(1)}_{\hat{3}}=2$ and $\Bar{h}^{(1)}_{\hat{3}}=0$, we obtain only one possible tensor structure:
\begin{equation}
    \langle O_1(P_1,\Theta_1)O_2(P_2) \hat{O}_3(P_{\hat{3}},\Theta_3) \rangle = \frac{H_{1\hat{3}}^{11}N_{\hat{3},12}^{1}f_1(\xi_1,\xi_2,\chi)}{(P_{12})^{\frac{\Delta_1+\Delta_2-\Delta_3}{2}}(P_{1\hat{3}})^{\frac{\Delta_1+\hat{\Delta}_3-\Delta_2}{2}}(P_{2\hat{3}})^{\frac{\Delta_2+\hat{\Delta}_3-\Delta_1}{2}}}.
\end{equation}

\subsection{Defect-Defect-Bulk}
The three-point correlator involving two defect and one bulk operator is not interesting by itself as it does not yield a crossing relation. However, we encounter an additional invariant in this case,
\begin{equation}
\begin{split}
    &\tilde{N}_{\hat{a}a}^{(i)}=\frac{C_{\hat{a}}^{iAB}P_{(\hat{a}+1)A}P_{aB}}{(P_{a\hat{a}})^{1/2}(P_{\hat{a}(\hat{a}+1)})^{1/2}(P_{a(\hat{a}+1)})^{1/2}} .
\end{split}
\end{equation}
Combining this new invariant with the previously known ones we obtain the following list of invariants:
\begin{equation}
    \begin{split}
        &H_3^{(i,j)}, \\
        &H_{3\hat{1}}^{(i,j)}, H_{3\hat{2}}^{(i,j)}, \\
        &G_{3\hat{1}}^{(i,j)},G_{3\hat{2}}^{(i,j)},\tilde{G}_{3\hat{1}}^{(i,j)},\tilde{G}_{3\hat{2}}^{(i,j)}, \\
        &H_{\hat{1}\hat{2}}^{(i,j)},\tilde{H}_{\hat{1}\hat{2}}^{(i,j)}, \\
        &K_{3\hat{1}}^i,K_{3\hat{2}}^i, \\
        &\tilde{N}_{\hat{1}3}^{(i)} ,\tilde{N}_{\hat{2}3}^{(i)} .\\
    \end{split}
\end{equation}
Only one cross-ratio can be constructed out of two defect and one bulk operators,
\begin{equation}
    \zeta=\frac{(P_{\hat{1}}\bullet P_{\hat{2}})(P_3 \bullet P_3)}{(P_{\hat{1}}\bullet P_3)(P_{\hat{2}}\bullet P_3)} .
\end{equation}

\subsection{Defect-Defect-Defect}
The last ingredients for implementing three-point bootstrap are three-point correlators of defect local operators. For three defect operators it is impossible to construct a cross-ratio. In addition to invariants appearing in (\ref{invariantsDD}), an additional invariant can be constructed,
\begin{equation} \label{invariantsDDD}
    \Tilde{K}_{\hat{a}\hat{b}}^{(i)}=\frac{C_{\hat{a}}^{{(i)}AB}P_{\hat{a}+1,A}P_{\hat{b}B}}{(P_{\hat{a}(\hat{a}+1)})^{1/2}(P_{(\hat{a}+1)\hat{b}})^{1/2}(P_{\hat{a}\hat{b}})^{1/2}} \quad \hat{b}\neq\hat{a},\hat{a}+1 .
\end{equation}
Using $n_2=3$ in ($\ref{numberofstructures}$), we determine the total number of invariants to be 9: 
\begin{equation}
    H_{\hat{1}\hat{2}}^{(i,j)}, H_{\hat{2}\hat{3}}^{(i,j)}, H_{\hat{3}\hat{1}}^{(i,j)}, \Tilde{H}_{\hat{1}\hat{2}}^{(i,j)}, \Tilde{H}_{\hat{2}\hat{3}}^{(i,j)}, \Tilde{H}_{\hat{3}\hat{1}}^{(i,j)}, \Tilde{K}_{\hat{1}\hat{3}}^{(i)}, \Tilde{K}_{\hat{2}\hat{1}}^{(i)},\Tilde{K}_{\hat{3}\hat{2}}^{(i)} .
\end{equation}
We will consider an example of a three-point correlator with a 2-form, a vector, and a scalar.
\begin{center}
$\lambda_{\hat{1}} =$
\ytableausetup{centertableaux}
\begin{ytableau}
  \\ 
  \\
\end{ytableau}
\hspace{1cm}$\lambda_{\hat{2}} =$
\ytableausetup{centertableaux}
\begin{ytableau}
  \\ 
\end{ytableau}
\hspace{1cm}$\lambda_{\hat{3}} = \bullet$
\end{center}
When all the quantum numbers of the defect operators are parallel to the defect, it acts like a correlator in an ordinary CFT. We list the result below,
\begin{equation}
    \langle \hat{O}_{\hat{\Delta}_1}(P_{\hat{1}},\Theta_1)\hat{O}_{\hat{\Delta}_2}(P_{\hat{2}},\Theta_2)\hat{O}_{\hat{\Delta}_3}(P_{\hat{3}})\rangle = \hat{f}_{\hat{O}_1\hat{O}_2\hat{O}_3}\frac{ H_{\hat{1}\hat{2}}^{(1,1)} \Tilde{K}_{\hat{1}\hat{3}}^{(1)} }{P_{\hat{1}\hat{2}}^{\frac{\hat{\Delta}_1+\hat{\Delta}_2-\hat{\Delta}_3}{2}}P_{\hat{2}\hat{3}}^{\frac{\hat{\Delta}_2+\hat{\Delta}_3-\hat{\Delta}_1}{2}}P_{\hat{3}\hat{1}}^{\frac{\hat{\Delta}_3+\hat{\Delta}_1-\hat{\Delta}_2}{2}}} .
\end{equation}
We can impose symmetrization by applying $Z\partial_{\Theta}$ derivatives ($\ref{mixed101}$) to the above expression. However, it is redundant in this case as there is no symmetry in any of the operator representations and $\boldsymbol{\Theta}$-basis serves us fine. Since parallel quantum numbers behave as a regular CFT, our result matches with that of \cite{mixed1}. If all the spins and forms are in the direction orthogonal to the defect, only one invariant is possible: $\Tilde{H}_{\hat{1}\hat{2}}^{(i,j)}$. However with just this invariant it is impossible to construct a tensor structure for the given operators.
\begin{equation}
    \langle \hat{O}_{\hat{\Delta}_1}(P_{\hat{1}},\Phi_1)\hat{O}_{\hat{\Delta}_2}(P_{\hat{2}},\Phi_2)\hat{O}_{\hat{\Delta}_3}(P_{\hat{3}})\rangle = 0
\end{equation}
We obtain different results depending on whether the spin and forms are aligned parallel or orthogonal to the defects. Mixed symmetric correlator between operators carrying both (parallel and orthogonal) quantum numbers can be computed in a similar manner.

\section{\texorpdfstring{$n$}{Lg}-Point Correlators}\label{secn}
In this section, we will briefly comment on $n$-point ($n=n_1+n_2$) correlators involving $n_1$ bulk and $n_2$ defect operators. The three-point correlators exhausted all possible invariants. No additional invariant can appear for higher point correlators and all the tensor structures have to be constructed out of the previously known invariants. We list all the invariants down together with their number,
\begin{equation}\label{allinv}
    \begin{split}
        &H_{a}^{(i,j)} \rightarrow n_1 \quad \text{where $i\neq j$} \quad || \quad H_{a \hat{a}}^{(i,j)} \rightarrow n_1n_2 \\
        &G_{a\hat{a}}^{(i)} \rightarrow n_1n_2 \quad || \quad \tilde{G}_{a\hat{a}}^{(i,j)} \rightarrow n_1n_2 \quad || \quad K_{a\hat{a}}^{i} \rightarrow n_1n_2 \quad  \\
        &S_{ab}^{(i,j)} \rightarrow \frac{n_1(n_1-1)}{2} \: \text{where $a\neq b$ and } S_{(ab)}^{(i,j)}=S_{(ba)}^{(j,i)} \\
        &\bar{S}_{ab}^{(i,j)} \rightarrow \frac{n_1(n_1-1)}{2} \: \text{where $a\neq b$ and } \Bar{S}_{(ab)}^{(i,j)}=\Bar{S}_{(ba)}^{(j,i)} \\
        &K_{ab}^{(i)} \rightarrow n_1^2-n_1 \: \text{where $a\neq b$} \quad || \quad \bar{K}_{ab}^{(i)} \rightarrow n_1^2-n_1 \: \text{where $a\neq b$} \\
        & H_{\hat{a}\hat{b}}^{(i,j)} \rightarrow \frac{n_2(n_2-1)}{2} \: \text{where $\hat{a}\neq \hat{b}$} \quad || \quad \tilde{H}_{\hat{a}\hat{b}}^{(i,j)} \rightarrow \frac{n_2(n_2-1)}{2} \: \text{where $\hat{a}\neq \hat{b}$}  \\
        &N_{\hat{k},1b}^{(i)} \rightarrow n_2(n_1-1) \: \text{where $b\neq 1$} \\
        &\tilde{K}_{\hat{a}\hat{b}}^{(i)} \rightarrow n_2 (n_2-2) \: \text{where $\hat{b}\neq \hat{a},\hat{a}+1$} \\
        &\tilde{N}_{\hat{a}a}^{(i)} \rightarrow n_1n_2 .
    \end{split}
\end{equation}
Tensor structures for $n$-point correlators have to be constructed out of these invariants while respecting the homogeneity constraints. There is a slight subtlety involved with the last three invariants. $N_{\hat{c}1b}$, $\tilde{N}_{\hat{a}a}$ and $\tilde{K}_{\hat{a}\hat{b}}$ are all independent at the three-point level. However, they are not all independent for a higher-point correlator as $N_{\hat{c}1b}$ can be generated from $\tilde{N}_{\hat{a}a}$ ($\ref{Bident}$). Depending on $n_1$ and $n_2$, the independence of $N_{\hat{c}1b}$, $\tilde{N}_{\hat{a}a}$ and $\tilde{K}_{\hat{a}\hat{b}}$ varies. We list down the different cases and the independent invariants associated to those cases:
\begin{equation}\label{subtle}
    \begin{split}
        &n_1\geq2 \quad \& \quad n_2=1 \qquad \rightarrow N_{\hat{c}ab}\\
        &n_1\geq1 \quad \& \quad n_2=2 \qquad \rightarrow \tilde{N}_{\hat{a}a}\\
        &n_1=0 \quad \& \quad n_2\geq 3 \qquad \rightarrow \tilde{K}_{\hat{a}\hat{b}}\\
        &n_1\geq1 \quad \& \quad n_2\geq 3 \qquad \rightarrow \tilde{N}_{\hat{a}a},\tilde{K}_{\hat{a}\hat{b}} .\\
    \end{split}
\end{equation}
In all other cases, $N_{\hat{c}ab}$, $\tilde{N}_{\hat{a}a}$ and $\tilde{K}_{\hat{a}\hat{b}}$ do not appear. We quote the equation for the total number of invariants ($\ref{numberofstructures}$) again for convenience, \begin{equation}\label{master2}
    3n_1^2-2n_1+2n_2^2-3n_2+5n_1n_2 .
\end{equation}
Taking into account ($\ref{subtle}$), the sum of all invariants listed in ($\ref{allinv}$) becomes,
\begin{equation}\label{limit}
    \begin{split}
        &n_1\geq2 \quad \& \quad n_2=1 \qquad \rightarrow 3n_1^2-2n_1+n_2^2-2n_2+5n_1n_2\\
        &n_1\geq1 \quad \& \quad n_2=2 \qquad \rightarrow 3n_1^2-2n_1+n_2^2-n_2+5n_1n_2\\
        &n_1=0 \quad \& \quad n_2\geq 3 \qquad \rightarrow 3n_1^2-2n_1+2n_2^2-3n_2+4n_1n_2\\
        &n_1\geq1 \quad \& \quad n_2\geq 3 \qquad \rightarrow 3n_1^2-2n_1+2n_2^2-3n_2+5n_1n_2 .\\
    \end{split}
\end{equation}
Except for the last case, this result does not seem to match with ($\ref{master2}$). The deviation from (\ref{master2}) can be calculated by subtracting our result from ($\ref{master2}$),
\begin{equation}
    \begin{split}
        &n_1\geq2 \quad \& \quad n_2=1 \qquad \rightarrow n_2^2-n_2=0 \implies n_2=1\\
        &n_1\geq1 \quad \& \quad n_2=2 \qquad \rightarrow n_2^2-2n_2=0 \implies n_2=2,0\\
        &n_1=0 \quad \& \quad n_2\geq 3  \qquad \rightarrow n_1n_2=0 \implies n_1=0\\
        &n_1\geq1 \quad \& \quad n_2\geq 3  \qquad \rightarrow 0  .\\
    \end{split}
\end{equation}
We see that all the polynomials in ($\ref{limit}$) yield the same result as ($\ref{master2}$) as they are various limits of the same equation ($\ref{master2}$) at different $n_1$ and $n_2$.

The number of independent cross-ratios for $n_1$ bulk and $n_2$ defect operators were calculated in \cite{radial} and the number is,
\begin{equation}
    n_1(n_1+1)+n_2(n_1+1) + \frac{n_2(n_2+1)}{2} .
\end{equation}
For the purpose of bootstrap in a defect CFT, higher (greater than three)-point correlators provide no new information. All of the defect CFT data is already accounted at three-point crossing level. 

For a purely bulk $n$-point correlator, tensor structures can be constructed out of the invariants in (\ref{invariantsBB}). When we count the total number of independent invariants keeping in mind each invariant has $i,j,...$ indices labelling the columns of Young representation as well, we get 
\begin{equation}
    \frac{1}{2}\bigg(\sum_a^nl^{(1)}_a\bigg)\bigg(\sum_a^n l^{(1)}_a+2n-\frac{3}{2}\bigg)+\sum_{a>b}l^{(1)}_a l^{(1)}_b.
\end{equation}

\section{Parity Analysis}\label{podd}
In our analysis so far, we have restricted to parity-even structures. In this section, we will consider parity-odd tensor structures. Parity entails a flip in one of the spatial directions. This implies that any Lorentz contraction would always be parity invariant. The Levi-Civita tensor $\epsilon$ is required to construct a tensor structure that is parity-odd. The $\epsilon$-tensor with all its indices contracted gives a contribution from each direction. Hence, the structures made out of $\epsilon$ are always parity odd. For the bulk operators which transforms under $O(d+1,1)$ representation, the epsilon tensor is the full $d+2$ dimensional one. Let us consider a spin-$1$ operator in the presence of a co-dimension 2 defect:
\begin{equation}
    \langle O_{\Delta}(P,Z) \rangle = a_O \frac{\epsilon_{01\cdots p+2IJ}P^{I} Z^{J}}{(P\circ P)^{\frac{\Delta + 1}{2}}}.
\end{equation}
The spin-$1$ correlator was zero in the parity-even case while it is non-zero here with parity-odd structure. In a similar manner, one-point correlators of completely anti-symmetric tensors (or forms) which were previously vanishing are non-zero using parity-odd structures. The following one-point correlators are possible for forms in the presence of a $q$ co-dimension defect:
\begin{equation}
    \langle O_{(q-1)\text{-form}}(P,\Theta) \rangle = \frac{\epsilon_{01\cdots p+2 I_1 \cdots I_q}P^{I_1}\Theta^{I_2}\cdots\Theta^{I_q}}{(P\circ P)^{\frac{\Delta+1}{2}}},
\end{equation}
\begin{equation}
    \langle O_{(p+1)\text{-form}}(P,\Theta) \rangle = \frac{\epsilon_{A_1 \cdots A_{p+2}12\cdots q}P^{A_1}\Theta^{A_2}\cdots\Theta^{A_{p+2}}}{(P\circ P)^{\frac{\Delta+1}{2}}},
\end{equation}
\begin{equation}
\begin{split}
        &\qquad \langle O_{q\text{-form}}(P,\Theta) \rangle= \\ 
        &\frac{(P\circ P)(\epsilon_{01\cdots p+2I_1 \cdots I_q}\Theta^{I_1}\cdots\Theta^{I_q})-q(P\circ \Theta)(\epsilon_{01\cdots p+2I_1 \cdots I_q}P^{I_1}\Theta^{I_2}\cdots\Theta^{I_q})}{(P\circ P)^{(\Delta+2)/2}}
\end{split}
\end{equation}
\begin{equation}
\begin{split}
    &\qquad \langle O_{(p+2)\text{-form}}(P,\Theta) \rangle= \\ 
    &\frac{(P\bullet P)(\epsilon_{A_1 \cdots A_{p+2}12\cdots q}\Theta^{A_1}\cdots\Theta^{A_{p+2}})-(p+2)(P\bullet \Theta)(\epsilon_{A_1 \cdots A_{p+2}12\cdots q}P^{A_1}\Theta^{A_2}\cdots\Theta^{A_{p+2}})}{(P\circ P)^{(\Delta+2)/2}} .
\end{split}
\end{equation}
We find that $(q-1)$, $(q)$, $(p+1)$ and $(p)$-forms can have a non-zero one-point correlator in the presence of a $q$ co-dimension defect. Once again we get a check of the defect duality ($\ref{duality}$). A defect of co-dimension $d+2-q$ gives a non-zero value to the same forms as a $q$ co-dimension defect. The structure of the above one-point correlators imply,
\begin{equation} \label{ConservEqn}
    \partial^M D_M \langle O_{n\text{-form}}(P,\Theta) \rangle=0 ,
\end{equation}
trivially. We do not obtain any constraints on the scaling dimension of the bulk operator from the above equation. The case for a non-zero expectation value of $(q-1)$-form and $(p+1)$-form has a clear physical picture. A defect CFT could have a $p$-form gauge potential $A_p$ sitting on the defect:
\begin{equation}
    S_{CFT}=S^{\prime}+\int_{\mathcal{M}_p} A_p.
\end{equation}
Here $S^{\prime}$ refers to other terms in the CFT action and the gauge potential $A_p$ is integrated over the entire defect. In such cases, the $(p+1)$-form field strength $dA_p$ can have a non-zero expectation value. The Hodge dual of the field strength $*dA_p$ is a $(m-1)$-form and it would also have a non-zero expectation value. Equation (\ref{ConservEqn}) can be explained by the fact that $d^2A_p$ and $d*dA_p$ vanish trivially.

Similarly, it is possible to construct parity-odd tensor structures for defect local operators. Defect operators have two quantum numbers, one for the parallel group and one for the transverse. This implies the defect operators can be parity-odd with respect to either. This is implemented by considering two separate $\epsilon$-tensors.
\begin{equation}
    \epsilon_{AB\cdots p+2} \quad \text{and}\quad \epsilon_{IJ\cdots q}
\end{equation}
Tensor structures constructed out of these two $\epsilon$-tensors will be parity-odd.
\section{Components}
Embedding space also simplifies the computation of conformal blocks. We would like to be able to carry out the conformal bootstrap program for defects directly in embedding space following the program initiated in \cite{embeddingboot1,embeddingboot2}. For completeness, we briefly mention the strategy to project down to physical space ($d$-dimensions) the results of previous sections. Only projections in the presence of flat defects are considered in this section. For a detailed review of component calculations for both spherical and flat cases we point the reader to \cite{marco}. To recover indices from a polynomial expression, the expression needs to be acted on by \textit{component derivatives}. These derivatives are constructed to remove the auxiliary vectors while maintaining the required symmetry or anti-symmetry. It is important to note that the form of these derivatives is operator-representation dependent. The derivatives listed below only work with symmetric traceless operators and forms.
\begin{equation}\label{compderivative}
\begin{split}
&D_z^a = (\frac{p-2}{2}+z^b \frac{\partial}{\partial z^b})\frac{\partial}{\partial z^a}-\frac{1}{2}z_a\frac{\partial^2}{\partial z^b \partial z_b}, \\
&D_w^i = (\frac{q-2}{2}+w^j \frac{\partial}{\partial w^j})\frac{\partial}{\partial w^i}-\frac{1}{2}w_i\frac{\partial^2}{\partial w^j \partial w_j}, \\
&D_{\theta}^a=\frac{p-2}{2}\frac{\partial}{\partial \theta^a}+\theta^b \frac{\partial}{\partial \theta^b}\frac{\partial}{\partial \theta^a}, \\
&D_{\phi}^i=\frac{q-2}{2}\frac{\partial}{\partial \phi^i}+\phi^j \frac{\partial}{\partial \phi^j}\frac{\partial}{\partial \phi^i}. 
\end{split}
\end{equation}
We have used $(a,i)$ to label physical space directions parallel and orthogonal to the defect. Projections to the Poincar\'{e} section for bulk operator in the presence of a flat defect are:
\begin{equation}
    \begin{split}
        &Z^{A(i)}|_x=(0,2x^mz^{(i)}_m,x^a),\qquad Z^{I(i)}|_x=z^{(i)i}, \\
        &\quad \Theta^{A(i)}|_x=(0,2x^m\theta^{(i)}_m,x^a),\qquad \Theta^{I(i)}|_x=\theta^{(i)i}, \\
        &P^A|_x=(1,x^mx_m,x^a),\qquad P^I|_x=x^i .
    \end{split}
\end{equation}
While the projections to Poincar\'{e} section for a defect operator are:
\begin{equation}
    \begin{split}
        &Z^{A(i)}|_x=(0,2x^az^{(i)}_a,x^a),\quad Z^I|_x=0,\quad W^A|_x=0,\quad W^{I(i)}|_x=w^{(i)i}, \\ &\Theta^{A(i)}|_x=(0,2x^a\theta^{(i)}_a,x^a),\quad \Theta^I|_x=0,\quad  \Phi^A|_x=0,\quad \Phi^{I(i)}|_x=\phi^{(i)i},  \\
        &P^A|_x=(1,x^ax_a,x^a) \text{  and}\quad P^I|_x=0.
    \end{split}
\end{equation}
Using these results, we can project the contractions between different vectors in physical space:
\begin{equation}\label{contraction}
\begin{split}
& Z_1^{(i)}\bullet Z_2^{(j)} \rightarrow z_1^{(i)a} z_2^{(j)b}\eta_{ab},   \quad P_m \bullet Z_n^{(j)} = x_{mn}^a z_n^{a(j)}-x_{n}^i z_n^{i(j)}\\
&-2P_m \bullet P_n =|x^a_{mn}|^2+|x^i_m|^2+|x^i_n|^2 , \\
& \Theta_1^{(i)}\bullet \Theta_2^{(j)} \rightarrow \theta_1^{(i)a} \theta_2^{(j)b}\eta_{ab},   \quad P_m \bullet \Theta_n^{(j)} = x_{mn}^a \theta_n^{a(j)}-x_{n}^i \theta_n^{i(j)} \, , \\
\end{split}
\end{equation}
where $x_{mn}=x_m-x_n$. We are now in a position to list down the steps to implement component calculation:
\begin{enumerate}
	\item For a correlator in embedding space, all the coordinates must be projected to the Poincar\'{e} patch and dot products evaluated via ($\ref{contraction}$).
	\item Depending on the correlator required component derivatives ($\ref{compderivative}$) must be acted accordingly.
\end{enumerate}
As an example, we will obtain the physical space result for a bulk two-point correlator involving a 2-form and a vector. Our goal is to compute  $\langle O_1^{[ab]}(x_1) O_2^c(x_2) \rangle$ from ($\ref{exampleBB1}$). We will directly work in $\boldsymbol{\theta}$-basis for components as there is no symmetry in the correlator indices. To obtain the correct correlator, the terms in ($\ref{exampleBB1}$) containing $\theta_1^a \theta_1^b \theta_2^c$ are required. Only one tensor structure contains the required terms.
\begin{equation}
\bar{S}_{12}^{(1,1)}\bar{K}_{12}^{(1)}=-\frac{(\Theta_1 \bullet \Theta_2)(P_1\circ P_1)(P_1 \circ P_2)(P_2 \circ P_2)(P_2\bullet \Theta_1)}{(P_1\circ P_1)^{3/2}(P_2\circ P_2)^2}
\end{equation}

Projecting down to $d$-dimensions we obtain,
\begin{equation}
\bar{S}_{12}^{(1,1)}\bar{K}_{12}^{(1)}|=\frac{(\theta_1^e \theta_2^f \eta_{ef})(x_{21}^g\theta_1^h\eta_{gh}-\theta_1^ix_2^i)(x_1^ix_2^i)}{|x_1^i||x_2^i|^2} .
\end{equation}
The structure of the required correlator suggests the form of the derivatives to be  $D_{\theta_2}^c D_{\theta_1}^b D_{\theta_1}^a$. The antisymmetry in the indices $a,b$ is manifest due to anti-commutation among $\theta_1$s.
\begin{equation}
\begin{split}
&D_{\theta_2}^c D_{\theta_1}^b D_{\theta_1}^a (\theta_1^e \theta_2^f \eta_{ef})(x_{21}^g\theta_1^h\eta_{gh})\frac{x_1^ix_2^i}{|x_1^i||x_2^i|^2} \\
&=(\alpha^3-\alpha^2)(\eta^{ac}x_{21}^b-\eta^{bc}x_{21}^a)\frac{x_1^ix_2^i}{|x_1^i||x_2^i|^2} \qquad \text{where} \: \alpha =(\frac{p-2}{2})
\end{split}
\end{equation}
This result has the desired antisymmetry in $a$ and $b$. The full correlator in physical space is,
\begin{equation}
\langle O_1^{[ab]}(x_1) O_2^c(x_2) \rangle =(\alpha^3-\alpha^2) \frac{(\eta^{ac}x_{21}^b-\eta^{bc}x_{21}^a)(x_1^ix_2^i)}{|x_1^i|^{\Delta_1}|x_2^i|^{2+\Delta_2}}f_6(\xi_1,\xi_2) .
\end{equation}

Even though this procedure for obtaining components is universal and works for arbitrary representations, the form of the derivative operators ($\ref{compderivative}$) is quite complicated for representations involving multiple $Z$s or $\Theta$s per operator. In those complicated cases, the procedure for calculating components has been given in \cite{mixed2}. Our goal is to work in embedding space itself so we will not follow this path.

\section{Defects in Arbitrary Representation of \texorpdfstring{$SO(q)$}{Lg}}
In previous sections, we had considered defects transforming as singlets under the global $SO(q)$ group. In this section, we will consider correlators of operators in the presence of a defect transforming in arbitrary representations of $SO(q)$. The defect will also have indices (symmetric, anti-symmetric, or mixed symmetric). We will contract defect indices with a $\Theta$-basis anti-symmetric auxiliary vector $\chi^I$ while demanding that $\chi^I$ is transverse. Schematically this looks like:
\begin{equation}
D^q(P_{\alpha})_{I_1\cdots I_n}\chi^{I_1}\cdots\chi^{I_n} .
\end{equation}
 We have defined $\chi$s to be transverse by construction. $\chi$s have the following property,
\begin{equation}
\chi^{(i)}\circ \chi^{(j)} = 0 .
\end{equation}
For defects indices we use $Y$ as a $\boldsymbol{Z}$-basis vector of the orthogonal group. We will only consider parity-even tensor structures of one and two-point correlators. We give an analogous formula  ($\ref{numberofstructures}$) to count the number of invariants (ignoring the $i-index$ of defect and operators):
\begin{equation}\label{numberofstructures2} 
    3n_1^2+2n_2^2-2n_2+5n_1n_2 .
\end{equation}
Dipole moments can be considered as vector-defects in a quantum field theory. In a conformal theory, defect in arbitrary representations under $SO(q)$ can be constructed by integrating an operator in the same representation of $SO(q)$ over the entire hyperplane of the defect. Schematically this looks like:
\begin{equation}
    D^q(P_{\alpha},\chi)=\int_Y O(Y,\Phi)|_Y\:\: d^qY ,
\end{equation}
where $O(Y,\Phi)$ has support only on the hyperplane.
\subsection{One-Point Correlator}
The new invariants that can appear in a one-point correlator of a bulk operator are the following,
\begin{equation}\label{spindefectinv1}
\begin{split}
\mathcal{P}_a^{(i)}=&\frac{P_a \circ \chi^{(i)}}{(P_{aa})^{1/2}},\\
\mathcal{R}_{a}^{(i,j)}=&\frac{C_a^{AI (i)}P_a^A\chi^{(j)}}{(P_{aa})} .
\end{split}
\end{equation}
We obtain the following invariants (including the previously known invariants),
\begin{equation}
H_1^{(i,j)},\mathcal{R}_{1}^{(i,j)},\mathcal{P}_1^{(i)} .
\end{equation}
The singlet defect case had only one invariant ($\ref{onepointbasis}$), whereas now there are three. As an example, let us consider a vector in the presence of a one-form (or vector) defect. Only one tensor structure can be constructed,
\begin{equation}
\langle O(\Theta) \rangle_{D(\chi) } = \frac{\mathcal{R}_{1}^{11}}{(X\circ X)^{\Delta/2}} .
\end{equation}
It is interesting to find that the vector operator has a non-zero one-point correlator. In the singlet defect case the one-point correlator of the vector vanishes.
\subsection{Two-Point Correlators}
\subsubsection{Bulk-Bulk}
In addition to the invariants listed in ($\ref{invariantsBB}$) and ($\ref{spindefectinv1}$), it might also be possible to construct the following invariant:
\begin{equation}
    \mathcal{T}_{ab}^{(i,j,k)}=\frac{C_a^{(i)AI}C_b^{(j)AJ}P_{aI}\chi_{J}^{(k)}}{P_{aa}(P_{bb})^{1/2}} .
\end{equation}
However this is not independent and it can be related to previously known invariants:
\begin{equation}
    (P_1\circ P_1)(P_2\circ P_2)^{1/2}(P_1 \bullet P_2) \mathcal{T}_{12}=(C_2^{AI}P_1^A\chi^I)(C_1^{AI}P_2^AP_1^I)+ \frac{C_1^{AB}C_{2AB}}{2}(P_1\circ P_1)(P_2\circ \chi) .
\end{equation}
Combining all the invariants together, any tensor structure has to be constructed out of the following invariants:

\begin{equation}
\begin{split}
&H_1^{(i,j)}, H_2^{(i,j)}, S_{12}^{(i,j)},\\
&\Bar{S}_{12}^{(i,j)}, K_{12}^{(i)}, K_{21}^{(i)}, \Bar{K}_{12}^{(i)}, \Bar{K}_{21}^{(i)}, \\
&\mathcal{P}_1^{(i)},\mathcal{P}_2^{(i)}, \\
&\mathcal{R}_{1}^{(i,j)},\mathcal{R}_{2}^{(i,j)} .
\end{split}
\end{equation}

For a defect in symmetric traceless representation, we can again use the trick of replacing all $\chi^i$-vector with a single $Y$-auxiliary vector. Tensor structures can be constructed out of these invariants for two-point correlators by equating homogeneity of the bulk operators with that of the product of invariants.

\subsubsection{Defect-Defect}
Only one new invariant appears in this case,
\begin{equation}
\bar{\mathcal{R}}_{\hat{a}}^{(i,j)}=\Phi_{\hat{a}}^{(i)} \circ \chi^{(j)} .
\end{equation}
Including the previously known invariants, the list of invariants in this case is:
\begin{equation}
H_{\hat{1}\hat{2}}^{(i,j)}, \tilde{H}_{\hat{1}\hat{2}}^{(i,j)},\bar{\mathcal{R}}_1^{(i,j)},\bar{\mathcal{R}}_2^{(i,j)} .
\end{equation}
If the defect operators only carry parallel quantum numbers, all correlators vanish. This is because the defect index is in the orthogonal direction and it needs another orthogonal index to contract with. The defect CFT becomes trivial in this case. It is necessary for defect local operators to carry orthogonal quantum numbers to have non-zero correlation functions in the case of defects with spin. 
\subsubsection{Bulk-Defect}
No new invariants can be constructed at this level. The possible invariants for a two-point correlator involving a bulk operators and a defect operator are,
\begin{equation}
\begin{split}
	&\mathcal{R}_{1}^{(i,j)},\mathcal{P}_1^i,\bar{\mathcal{R}}_{\hat{2}}^{(i,j)} \\
	&H_{1\hat{2}}^{(i,j)}, H_1^{i,j}, G_{1\hat{2}}^{i},\tilde{G}_{1\hat{2}}^{(i,j)}, K_{1\hat{2}}^{i} .
\end{split}
\end{equation}
As an example, we would like to know the bulk scalar decomposition in the presence of a defect transforming as a $m$-form under $SO(q)$. In this case the only invariants that we can use are $\mathcal{P}_1^i,G_{1\hat{2}}^{i},\bar{\mathcal{R}}_{\hat{2}}^{i}$. We find that only defect operators whose representation (under $SO(q)$) has a height less or equal to $m+1$ appear in the decomposition. When the $m$-form defect is a 0-form (singlet) the maximum height of defect operator-representation is one, the same as shown in ($\ref{scalardecomp}$).

\section{Concluding Remarks}

We have constructed correlators of operators in a theory where the symmetry group is broken into $SO(p+1,1) \times SO(q)$. We were able to identify the different representations of defect operators that can appear in the bulk-to-defect expansion of a given bulk operator. We have also computed all the invariants that can appear at the level of one-point, two-point and three-point correlators. Their generalizations to $n$-point correlators are also discussed. We also discuss one and two-point correlators for defects transforming in arbitrary representations of the orthogonal $SO(q)$ group.

With these results in hand it would be possible to constrain the defect CFT by studying crossing relation of operators in arbitrary representations. A defect CFT (dCFT) has two sets of CFT data in addition to couplings between the bulk and the defect sector. The total data-set of a dCFT is:
\begin{equation} \label{dcftdata}
\{\Delta,\hat{\Delta},f_{ijk},\hat{f}_{ijk},b_{ij}\} .
\end{equation}

The four-point crossing equation for the theory living on the defect (in principle) fixes all the data of the defect sector. The remaining information about the bulk and the bulk-to-defect couplings are captured by crossing equations of the $\langle O_1 O_2 \rangle$ and $\langle O_1 O_2 \hat{O}_3 \rangle$ correlators. As an example let us consider a two-point correlator of two bulk scalars.
The bulk two point function has two expansion channels (Figure 3), U and Y \footnote{We thank Daniel Robbins for the terminology.} . They yield a crossing equation in terms of $\{b_{ij},f_{ijk}\}$,
\begin{figure}\label{uychannel}
	\centering
	\includegraphics[scale=0.7]{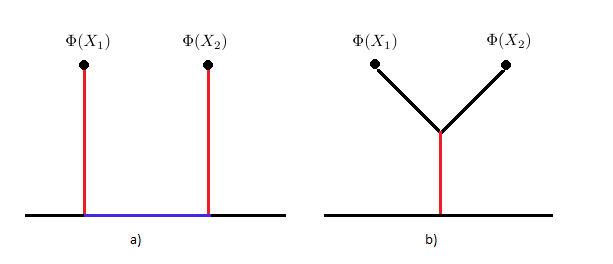}
	\caption{a) U-Channel: The bulk operators are decomposed in terms of defect operators. We then take a two-point correlator of the resulting defect operators. b) Y-Channel: First we take an OPE of two bulk operators and then decompose the resulting operator in terms of defect local operator.}
	\label{fusionfig}
\end{figure}

\begin{equation}\label{defectcrossing}
\sum_{\hat{O}}b^2_{\Phi \hat{O}}F(\hat{\Delta}_{\hat{O}},\eta)=\sum_{O}f_{\Phi \Phi O}b_{O1}\tilde{F}(\Delta_O,\eta) .
\end{equation}
$F$ and $\tilde{F}$ are conformal blocks which are functions of scaling dimensions and relevant cross ratios. Their explicit form was calculated in \cite{marco,osborn}. The crossing equation has been studied both analytically \cite{marco,marco3} and numerically \cite{marco2,boundaryboots}. This problem is challenging to solve numerically as the right-side ($\ref{defectcrossing}$) does not have positive coefficients. This crossing relation does not provide us with the complete information of the dCFT data as we are missing $\hat{f}_{ijk}$. To constrain the remaining data we need crossing arising from three-point correlator involving two bulk and one defect local operator. We hope that calculations done in this paper would come in handy for three-point bootstrap. A three-point crossing involving scalars has the following schematic form:
\begin{equation}
  \sum_{\hat{O}_1}^{}\sum_{\hat{O}_2}^{}b_{\Phi_1 \hat{O}_1}b_{\Phi_2 \hat{O}_2}\hat{f}_{\hat{O}_1 \hat{O}_2 \hat{O}_3}\tilde{G}(\eta,\hat{\Delta}_1,\hat{\Delta}_2) =  \sum_{\tilde{\Phi}}f_{\Phi_1 \Phi_2 \tilde{\Phi}}b_{\tilde{\Phi} \hat{O}_3}G(\Delta_{\tilde{\Phi}},\eta) .  
\end{equation}
$\Delta$ and $\hat{\Delta}$ stand for scaling dimensions of operators appearing in the intermediate channels. $\hat{G}$ and $\tilde{G}$ are the conformal blocks, which are functions of cross-ratios. These functions can be determined by acting with the Casimir operator as was done in \cite{marco}. These blocks were recently calculated in \cite{defectmanifolds} for the boundary case. In a future work we would like to calculate the conformal blocks for three-point correlators for arbitrary co-dimension defects.

 Tensor structures calculated in this work would also be relevant for lightcone analytic bootstrap of defect CFTs \cite{lightcone}. Fusion of two or more defects should also be an interesting problem to tackle in higher dimensions. It would also be interesting to apply the shadow formalism of conformal block calculation \cite{shadow} in the defect case. Analytic continuation of defect CFTs to the Lorentzian signature is also interesting. In this case, it would be interesting to study the expectation value of light-ray operators \cite{Kravchuk:2018htv} and consequently the average null energy condition (ANEC). It has recently been shown \cite{anec1,anec2} that the ANEC can be used to place a lower bound on operator dimensions. It would be interesting to apply this method to the case of a defect CFT. In a future work our goal is to report progress in these directions. 

\paragraph{Acknowledgments}
We wish to thank Andrew B. Royston, Daniel Robbins, Marco Meineri and William Linch III for extremely helpful discussions and correspondence. We also would like to thank Katrin Becker, Dmitry Ponomarev, Taylor Whitehead and Lauren Huff for constructive criticism on the manuscript.
\appendix
\section{Notations}\label{notations}
We will summarize the notation used throughout the paper in this section. Notations for directions are,
\begin{equation}
    \begin{split}
        &M,N,\cdots \rightarrow \text{Directions of the embedding space.}\\
        &A,B,\cdots \rightarrow \text{Directions parallel to the defect in the embedding space.}\\
        &I,J,\cdots \rightarrow \text{Directions orthogonal to the defect in the embedding space.} \\
        &m,n,\cdots \rightarrow \text{Directions in physical space.} \\
        &a,b,\cdots \rightarrow \text{Direction parallel to defect in physical space.} \\
        &i,j,\cdots \rightarrow \text{Directions othogonal to the defect in physical space.} \\
\end{split}
\end{equation}
Notation for position and auxiliary vectors:
\begin{equation}
    \begin{split}
      &P_a \rightarrow \text{Position of bulk local operator a.} \\
        &P_{\hat{a}} \rightarrow \text{Position of defect local operator $\hat{a}$.} \\
        &\Theta_{a}^{(i)}/Z_{a}^{(i)} \rightarrow \text{Auxiliary vector associated with i-th column/row of bulk operator.}\\
        &\Theta_{\hat{a}}^{(i)}/Z_{\hat{a}}^{(i)} \rightarrow \text{Auxiliary vector associated with i-th column/row of defect operator ($SO(p+1,1))$).}\\ 
        &\bar{\Theta}_{\hat{a}}^{(i)}/\bar{Z}_{\hat{a}}^{(i)} \rightarrow \text{Auxiliary vector associated with i-th column/row of defect operator ($SO(q)$).}\\ 
    \end{split}
\end{equation}
Notation for representation:
\begin{equation}
    \begin{split}
        &n^{C/R}_a \rightarrow \text{Number of columns/rows in bulk-operator a.} \\
        &n^{C/R}_{\hat{a}} \rightarrow \text{Number of columns/rows in defect-operator $\hat{a}$. } \\
        &n^{C/R}_{\hat{a}} \rightarrow \text{Number of columns/rows in defect-operator $\hat{a}$. } \\
        &\lambda_a \rightarrow \text{Representation of a bulk operator under $SO(d+1,1)$.}\\
        &\lambda_{\hat{a}} \rightarrow \text{Representation of a defect operator under $SO(1+p,1)$.}\\
        &\bar{\lambda}_{\hat{a}} \rightarrow \text{Representation of a defect operator under $SO(q)$.}\\
        &l^{(i)}_{a}/h^{(i)}_{a} \rightarrow \text{Length/height of i-th row/column of bulk operator.}\\
        &l^{(i)}_{\hat{a}}/h^{(i)}_{\hat{a}} \rightarrow \text{Length/height of i-th row/column of defect operator under $SO(p+1,1)$.}\\
        &\bar{l}^{(i)}_{\hat{a}}/\bar{h}^{(i)}_{\hat{a}} \rightarrow \text{Length/height of i-th row/column of defect operator under $SO(q)$.}\\
     \end{split}
\end{equation}
Notation of operators and couplings:
\begin{equation}
    \begin{split}
        &O \rightarrow \text{Bulk operator.} \\
        &\hat{O} \rightarrow \text{Defect operator.} \\
        &b_{O \hat{O}} \rightarrow \text{Bulk-to-defect coupling between bulk $O$ and defect $\hat{O}$.} \\
        &f_{OOO} \rightarrow \text{Three-point coupling of Bulk sector.} \\
        &\hat{f}_{\hat{O}\hat{O}\hat{O}} \text{Three-point coupling of defect sector.}\\
        &\Delta \rightarrow \text{Scaling dimension of bulk operator.} \\
        &\hat{\Delta} \rightarrow \text{Scaling dimension of defect operator.}\\
    \end{split}
\end{equation}
\section{Invariants} \label{c-contra}
We will list down all invariants schematically (and suppressing the $i$-indices) beginning with no C-tensor case. Hats on vectors means that they are associated with defect local operators. We consider both $SO(1+p,1)$ and $SO(q)$ contractions together:
\begin{equation}
    \begin{split}
        &P\Phi \rightarrow G_{a\hat{a}},\\
        &\Phi\Phi \rightarrow \tilde{H}_{\hat{a}\hat{b}}.\\
    \end{split}
\end{equation}
Moving on to single C-tensor case:
\begin{equation}
    \begin{split}
        &CPP \rightarrow K_{ab},\bar{K}_{ab},\\
        &CP\hat{P} \rightarrow K_{a\hat{a}},\\
        &C\hat{P}\hat{P} \rightarrow \text{can be reduced using $K_{a\hat{a}}$},\\
        &CP\Phi \rightarrow \tilde{G}_{a\hat{a}},\qquad C\hat{P}\Phi \rightarrow  \text{can be reduced using $\tilde{G}_{a\hat{a}}$ and $K_{a\hat{a}}$ },\\
        &\hat{C}PP \rightarrow N_{\hat{k}ab},\\
        &\hat{C}P\hat{P} \rightarrow \tilde{N}_{\hat{a}a},\\
        &\hat{C}\hat{P}\hat{P} \rightarrow \tilde{K}_{\hat{a} \hat{b}},\\
        &\hat{C}P\Phi \rightarrow \text{not possible},\qquad C\hat{P}\Phi \rightarrow \text{not possible} .\\
    \end{split}
\end{equation}
Moving on to two bulk C-tensor contractions:
\begin{equation}
    \begin{split}
        &CC \rightarrow H_a, \\
        &CCPP \rightarrow S_{ab},\bar{S}_{ab},\\
        &CCP\hat{P} \rightarrow \text{can be reduced using $S_{ab},\bar{S}_{ab}$},\\
        &CC\hat{P}\hat{P} \rightarrow \text{can be reduced using $S_{ab},\bar{S}_{ab}$},\\
        &CCP\Phi \rightarrow \text{can be reduced using $CP\Phi$ and $H_{ab}$}, \qquad CC\hat{P}\Phi \rightarrow \text{can be reduced}  .
    \end{split}
\end{equation}
Two defect C-tensor contractions:
\begin{equation}
    \begin{split}
        &\hat{C}\hat{C} \rightarrow  H_{\hat{a}\hat{b}},\\
        &\hat{C}\hat{C}PP \rightarrow \text{can be reduced using $H_{\hat{a}\hat{b}}$},\\
        &\hat{C}\hat{C}P\hat{P} \rightarrow \text{can be reduced},\\
        &\hat{C}\hat{C}\hat{P}\hat{P} \rightarrow \text{can be reduced},\\
        &\hat{C}\hat{C}P\Phi \rightarrow \text{not possible},\qquad \hat{C}\hat{C}\hat{P}\Phi \rightarrow  \text{not possible} .
    \end{split}
\end{equation}
Lastly we consider one defect C-tensor and one bulk C-tensor contractions,
\begin{equation}
    \begin{split}
        &\hat{C}C \rightarrow H_{a \hat{a}}, \\
        &\hat{C}CPP \rightarrow \text{can be reduced using $H_{a \hat{a}}$},\\
        &\hat{C}CP\hat{P} \rightarrow \text{can be reduced}, \\
        &\hat{C}C\hat{P}\hat{P} \rightarrow \text{can be reduced},\\
        &\hat{C}CP\Phi \rightarrow \text{can be reduced},\qquad \hat{C}C\hat{P}\Phi \rightarrow \text{can be reduced} .
    \end{split}
\end{equation}

\section{Useful Identities} \label{Bident}
In this section we will list down some important identities involving C-tensors. We will first begin with single C-tensor case:
\begin{equation}
    (P_1\circ P_2)C_2^{AI}P_{1A}P_{2I}=(P_1\bullet P_2)C_2^{AI}P_{2A}P_{1I}+(P_2\circ  P_2)C_2^{AI}P_{1A}P_{1I},
\end{equation}
\begin{equation}
    \begin{split}
    C_1^{AI}P_{2A}P_{3I}=&\frac{(P_1 \bullet P_2)}{(P_1 \bullet P_1)}C_1^{AI}P_{1A}P_{3I}+\frac{(P_1 \circ P_3)}{(P_1 \circ P_2)}C_1^{AI}P_{2A}P_{2I} \\
    &-\frac{(P_1 \circ P_3)(P_1\bullet P_2)}{(P_1 \circ P_2)(P_1 \bullet P_1)}C_1^{AI}P_{1A}P_{2I} .
    \end{split}
\end{equation}
Moving on to two C-tensor identities,
\begin{equation}
    (P_1 \circ P_1)C_1^{AI}C_2^{AJ}P_{3I}\Phi_J=(P_1\circ P_3)C_1^{AI}C_2^{AJ}P_{1I}\Phi_{J}-(C_2^{AI}P_{1A}\Phi_{J})(C_1^{AI}P_{1A}P_{3I}),
\end{equation}
\begin{equation}
\begin{split}
    C_1^{AI}C_2^{AJ}P_{2I}P_{1J}=&\frac{(P_1\circ P_2)^2}{(P_1\circ P_1)(P_2\circ P_2)}C_1^{AI}C_2^{AJ}P_{1I}P_{2J}-\frac{(P_1\bullet P_2)}{(P_1\circ P_1)(P_2\circ P_2)}C_1^{AI}P_{1A}P_{2I}C_2^{BJ}P_{2B}P_{1J}\\
    &-\frac{1}{(P_1\circ P_1)}C_1^{AI}P_{1A}P_{2I}C_2^{BJ}P_{1B}P_{1J}-\frac{1}{(P_2\circ P_2)}C_1^{AI}P_{2A}P_{2I}C_2^{BJ}P_{2B}P_{1J},
\end{split}    
\end{equation}
\begin{equation}
    (P_2 \bullet P_2)C_1^{AI}C_2^{BI}P_{1I}P_{3J}=(P_2\bullet P_3)C_1^{AI}C_2^{BI}P_{1A}P_{2B}+(C_1^{AI}P_{1A}P_{2I})(C^{AB}_2P_{3A}P_{2B}) .
\end{equation}

\section{Equation for Tensor Structures}
In this section we will list down the non-negative integer equations for different correlators.
\subsection{\texorpdfstring{$\langle O O \rangle$}{Lg}}\label{OOappendix}
We list down the powers of different invariants in a tensor structure,
\begin{equation}
    \begin{split}
        & H_1^{(i,j)}\rightarrow a_{ij}, \quad H_2^{(i,j)}\rightarrow b_{ij},\quad S_{12}^{(i,j)}\rightarrow c_{ij},\quad \Bar{S}_{12}^{(i,j)} \rightarrow d_{ij}, \quad K_{12}^{(i)} \rightarrow e_i, \\
        &K_{21}^{(i)}\rightarrow f_i, \quad \Bar{K}_{12}^{(i)}\rightarrow g_i, \quad \Bar{K}_{21}^{(i)} \rightarrow h_i .
    \end{split}
\end{equation}
Using a similar notation listed in ($\ref{diph1}$) we have:
\begin{equation}
    \begin{split}
        &h^{(i)}_{1}=\sum_j^{n^C_1}a_{ij}+\sum_j^{n^C_2}c_{ij}+\sum_j^{n^C_2}d_{ij}+e_{i}+g_{i}\\
        &h^{(i)}_{2}=\sum_j^{n^C_1}b_{ij}+\sum_j^{n^C_1}c_{ji}+\sum_j^{n^C_1}d_{ji}+f_{i}+h_{i}        \end{split}
\end{equation}
\subsection{\texorpdfstring{$\langle OOO \rangle$}{Lg}}\label{OOOappendix}
Powers of each invarinat are denoted as,
\begin{equation}
\begin{split}
    &H_1^{(i,j)} \rightarrow a_{ij},\quad  H_2^{(i,j)} \rightarrow b_{ij}, \quad H_3^{(i,j)} \rightarrow c_{ij}, \\
    &S_{12}^{(i,j)} \rightarrow d_{ij}, \quad S_{23}^{(i,j)} \rightarrow e_{ij}, \quad  S_{31}^{(i,j)} \rightarrow f_{ij}, \\
    &\Bar{S}_{12}^{(i,j)} \rightarrow g_{ij}, \quad \Bar{S}_{23}^{(i,j)} \rightarrow h_{ij}, \quad \Bar{S}_{31}^{(i,j)} \rightarrow i_{ij}, \\
    &K_{12}^{(i)}\rightarrow j_{i}, \quad K_{21}^{(i)} \rightarrow k_{i}, \quad K_{23}^{(i)} \rightarrow l_{i}, \quad K_{32}^{(i)} \rightarrow m_{i}, \quad K_{31}^{(i)} \rightarrow n_{i}, \quad K_{13}^{(i)} \rightarrow o_{i},\\ &\Bar{K}_{12}^{(i)}\rightarrow p_{i}, \quad \Bar{K}_{21}^{(i)} \rightarrow q_{i}, \quad \Bar{K}_{23}^{(i)}\rightarrow r_{i}, \quad  \Bar{K}_{32}^{(i)}\rightarrow s_{i}, \quad \Bar{K}_{31}^{(i)} \rightarrow t_{i}, \quad \Bar{K}_{13}^{(i)} \rightarrow u_{i} .
\end{split}
\end{equation}
We get the following system of equation:
\begin{equation}
    \begin{split}
        &h^{(i)}_{1}=\sum_j^{n^C_1}a_{ij}+\sum_j^{n^C_2}d_{ij}+\sum_j^{n^C_3}f_{ji}+\sum_j^{n^C_2}g_{ij}+\sum_j^{n^C_3}i_{ji}+j_{i}+o_{i}+p_i+u_i,\\
        &h^{(i)}_{2}=\sum_j^{n^C_2}b_{ij}+\sum_j^{n^C_1}d_{ji}+\sum_j^{n^C_3}e_{ij}+\sum_j^{n^C_1}g_{ji}+\sum_j^{n^C_3}h_{ij}+k_{i}+l_{i}+q_i+r_i,\\
        &h^{(i)}_{3}=\sum_j^{n^C_3}c_{ij}+\sum_j^{n^C_2}e_{ji}+\sum_j^{n^C_1}f_{ij}+\sum_j^{n^C_2}h_{ji}+\sum_j^{n^C_1}i_{ij}+m_{i}+n_{i}+s_i+t_i .
    \end{split}
\end{equation}
\subsection{\texorpdfstring{$\langle O O\hat{O}\rangle$}{Lg}}\label{OOhatOappendix}
Let us denote the power of each invariant by the following symbols,
\begin{equation}
    \begin{split}
        &G^{(i)}_{1\hat{3}}\rightarrow a_i, \quad G^{(i)}_{2\hat{3}}\rightarrow b_i, \quad H_{1}^{(i,j)}\rightarrow c_{ij}, \quad H_{2}^{(i,j)}\rightarrow d_{ij}, \quad \tilde{K}_{12}^{(i)} \rightarrow e_i, \quad K_{12}^{(i)} \rightarrow f_i \\
        &\tilde{K}_{21}^{(i)} \rightarrow g_i, \quad K_{21}^{(i)} \rightarrow h_i, \quad \tilde{G}_{1\hat{3}}^{(i,j)}\rightarrow i_{ij}, \quad \tilde{G}_{2\hat{3}}^{(i,j)}\rightarrow j_{ij},\quad H_{1\hat{3}}^{(i,j)}\rightarrow k_{ij}, \quad H_{2\hat{3}}^{(i,j)}\rightarrow l_{ij}, \\
        &S_{12}^{(i,j)}\rightarrow m_{ij},\quad \bar{S}_{12}^{(i,j)}\rightarrow n_{ij}, \quad K_{1\hat{3}}^{(i)} \rightarrow o_i ,\quad K_{2\hat{3}}^{(i)} \rightarrow p_i,\quad N_{\hat{3},12}^{(i)} \rightarrow q_i .
    \end{split}
\end{equation}
Let the number of $\boldsymbol{\Theta}$-rows of the bulk operators be $n^C_1$ and $n^C_1$. For the defect operator we have two quantum number whose $\Theta$ and $\Phi$ rows are $n^C_{\hat{3}}$ and $\bar{n}^C_{\hat{3}}$,
\begin{equation}
    \begin{split}
        &h^{(i)}_{1}=\sum_{j}^{n^C_1}c_{ij}+e_i+f_i+\sum_j^{\bar{n}^C_{\hat{3}}}i_{ij}+\sum_{j}^{n^C_{\hat{3}}}k_{ij}+\sum_j^{n^C_2}m_{ij}+\sum_j^{n^C_2}n_{ij}+o_i, \\
        &h^{(i)}_{2}=\sum_{j}^{n^C_2}d_{ij}+g_i+h_i+\sum_j^{\bar{n}^C_{\hat{3}}}j_{ij}+\sum_{j}^{n^C_{\hat{3}}}l_{ij}+\sum_j^{n^C_1}m_{ji}+\sum_j^{n^C_1}n_{ji}+p_i, \\
        &h^{(i)}_{\hat{3}}=\sum_{j}^{n^C_1}k_{ji}+\sum_{j}^{n^C_2}l_{ji}+q_i, \\
        &\bar{h}^{(i)}_{\hat{3}}=a_i+b_i+\sum_{j}^{n^C_1}i_{ji}+\sum_{j}^{n^C_2}j_{ji} .
    \end{split}
\end{equation}


\end{document}